\documentclass[10pt, twocolumn, twoside]{IEEEtran}
\IEEEoverridecommandlockouts

\ifCLASSINFOpdf
 \usepackage[pdftex]{graphicx}
\else
 \usepackage[dvips]{graphicx}
\fi
\usepackage{cite}
\usepackage{url}
\usepackage{amsmath,amssymb,amsfonts, lipsum}
\usepackage{float}
\usepackage{algorithmic}
\usepackage{epstopdf}
\usepackage{graphicx}
\usepackage{textcomp}
\usepackage{color}
\usepackage{xcolor}
\usepackage{multirow}
\usepackage{algorithm}
\usepackage{stfloats}
\usepackage{makecell}
\usepackage{times}
\usepackage[caption=false, font=footnotesize, labelfont=rm, textfont=rm]{subfig}
\usepackage{booktabs}
\usepackage{array}
\usepackage{tabularx}
\usepackage{threeparttable}

\makeatletter
\let\myorg@bibitem\bibitem
\def\bibitem#1#2\par{
  \@ifundefined{bibitem@#1}{
    \myorg@bibitem{#1}#2\par
  }{
    \begingroup
      \color{\csname bibitem@#1\endcsname}
      \myorg@bibitem{#1}#2\par
    \endgroup
  }
}
\makeatother

\def\BibTeX{{\rm B\kern-.05em{\sc i\kern-.025em b}\kern-.08em
  T\kern-.1667em\lower.7ex\hbox{E}\kern-.125emX}}

\begin{document}
\title{R$^{2}$Net: 2D Deep Residual Learning with Height Embedding for 3D Radio Map Estimation}

 \author{\IEEEauthorblockN{Huiting~Rao, Junyuan~Wang,~\IEEEmembership{Member,~IEEE,} Huiling~Zhu,~\IEEEmembership{Senior Member,~IEEE,} and Cheng-Xiang~Wang,~\IEEEmembership{Fellow,~IEEE}
        \thanks{Copyright \copyright\ 2026 IEEE. Personal use of this material is permitted. However, permission to use this material for any other purposes must be obtained from the IEEE by sending a request to pubs-permissions@ieee.org.}
        \thanks{Received 3 March 2025; revised 14 November 2025 and 3 March 2026; accepted 27 April 2026. This work was supported in part by the National Natural Science Foundation of China under Grant 62371344 and in part by the Fundamental Research Funds for Central Universities. The work of Cheng-Xiang Wang was support in part by the Major Science and Technology Project of Jiangsu Province under Grant BG2025039 and in part by the Research Fund of National Mobile Communications Research Laboratory, Southeast University, under Grant 2026A05. An earlier version of this paper was presented in part at the IEEE International Conference on Communications (ICC), Denver, United States, in June 2024 \cite{R2NET}. The associate editor coordinating the review of this article and approving it for publication was X. Yuan. \emph{(Corresponding author: Junyuan Wang).}}
        \thanks{Huiting Rao is with the College of Electronic and Information Engineering, Tongji University, Shanghai 201804, China (email: huiting\_rao@tongji.edu.cn).}
        \thanks{Junyuan Wang is with the College of Electronic and Information Engineering, Shanghai Institute of Intelligent Science and Technology, and the Institute of Advanced Study, Tongji University, Shanghai 201804, China (e-mail: junyuanwang@tongji.edu.cn).}
        \thanks{Huiling Zhu is with the School of Engineering, University of Kent, CT2 7NT Canterbury, U.K. (email: h.zhu@kent.ac.uk).}
        \thanks{Cheng-Xiang Wang is with the National Mobile Communications Research Laboratory, School of Information Science and Engineering, Southeast University, Nanjing 210096, China, and also with Purple Mountain Laboratories, Nanjing 211111, China (email: chxwang@seu.edu.cn).}}}

\maketitle

\begin{abstract}
Acquiring channel knowledge is required by many applications. For instance, handover in cellular networks is mainly decided based on the knowledge of pathloss. In contrast to traditional statistical distance-determined models that might provide misleading pathloss estimates, researchers started to explore deep learning methods recently to accurately estimate the radio map that characterizes the spatial distribution of pathloss according to the specific physical wireless propagation environment. However, existing works mainly focused on 2D radio map estimation by assuming that all receivers are at the same height. In fact, radio maps could be significantly different at different receiver heights, highlighting the importance of 3D radio map estimation. In this paper, we first propose a method to embed height information into 2D images, and then propose a general 2D radio residual network (R$^{2}$Net) for 3D radio map estimation. Since pathloss exhibits different characteristics in indoor and outdoor scenarios, we specifically propose R$^{2}$Net-In for indoor scenarios and R$^{2}$Net-Out for outdoor scenarios to better capture penetration loss and diffraction loss, respectively. Extensive experimental results show that our R$^{2}$Net significantly outperforms the state-of-the-art benchmarks in terms of estimation accuracy, computational and storage costs, and inference speed. In addition, due to the lack of publicly available 3D radio map datasets, a 3D indoor radio map dataset (3DiRM3200) is created, which took more than $1,000$ labour hours. The dataset and codes will be available at \url{https://github.com/lighttime2023/3DiRM3200.git}.
\end{abstract}

\begin{IEEEkeywords}
3D radio map estimation, pathloss prediction, height embedding, computer vision, deep learning
\end{IEEEkeywords}

\section{Introduction}
Estimating the received signal strength (RSS) has always been a crucial task in wireless communications. A radio map characterizes the spatial distribution of the RSS at different locations in the form of an image, where each pixel represents the corresponding pathloss. Radio maps have showcased wide applications in many areas, such as user handover, link scheduling, aerial base-station (BS) placement, spectrum sharing, fingerprint-based localization and digital twins~\cite{Spectrum-Sharing,ABS-Placement,Robot-Path-Planning,AINetworking,6GVisions,CKMTutorial,LocUNet}. For instance, with the help of radio maps, the accuracy and efficiency of fingerprint-based localization can be significantly improved by exploring deep learning methods~\cite{LocUNet}. Due to the importance to many applications, radio map estimation is gaining increasing momentum recently, especially with the rapid development of deep learning techniques.

The conventional way to construct a radio map is to collect pathloss measurements manually via site survey. Yet, there are usually some areas that are not physically accessible, leading to missing data on the radio map. Radio map reconstruction~\cite{reconstruction_sun,Bayesian,Spectrummap} was thus studied to predict the unmeasured pathloss or RSS based on the measured data at other locations, which are either pure data-driven or data-model dual-driven. For instance, a pure data-driven matrix completion method was employed in~\cite{reconstruction_sun} to exploit the potential low rank structure of the radio map, while data-model dual-driven strategies were proposed in~\cite{Bayesian} and~\cite{Spectrummap} to estimate the missing RSS data on the radio map from sparse measurements with the help of radio propagation models. The above site survey and radio map reconstruction methods work only in the scenarios where BSs have been deployed. However, radio maps are also essential in the network planning phase, such as for BS placement optimization. Simulation software, e.g., WinProp~\cite{WinProp}, has been developed to construct the radio map for a specific physical environment by searching for and simulating possible propagation paths between the transmitter and the receiver. However, these simulation methods could be very time-consuming, especially when they are used in a complex environment with rich signal reflections, diffractions, etc.

Recently, thanks to the great potentials shown by machine learning, learning based methods have been explored for radio map estimation~\cite{KNN,Gaussian-Regression,Gaussian-Process,RadioUNet,FadeNet,Transformerradionet,PPNet,2D_to_3D,Pano2RSSI,RadioResUNet,IndoorRSSINet,two-rooms_fixed_transmitter}, which can be classified as traditional machine learning methods~\cite{KNN,Gaussian-Regression,Gaussian-Process} or deep learning based computer vision methods~\cite{RadioUNet,FadeNet,Transformerradionet,PPNet,2D_to_3D,Pano2RSSI,IndoorRSSINet,RadioResUNet,two-rooms_fixed_transmitter}. Similar to channel modeling practice~\cite{PervasiveChannel,6GPervasiveChannel}, the traditional machine learning methods usually modeled pathloss as a function of receiver location and aimed to learn the relationship between the receiver position and the RSS~\cite{KNN,Gaussian-Regression,Gaussian-Process}. As a result, to predict the radio maps for different transmitter locations or in different propagation environments, different models need to be trained. By contrast, for the deep learning based computer vision methods, once the model is successfully trained, they can estimate the radio map for any propagation environment with any transmitter location, which is more efficient.

Existing deep learning based computer vision methods for radio map estimation were mainly developed for outdoor scenarios~\cite{RadioUNet,FadeNet,Transformerradionet,PPNet,2D_to_3D}, where BSs are usually equipped at much higher altitude than ground users, such as at the roofs of buildings. In such cases, the propagation environment was approximately the same for different ground users at different heights, i.e., the radio map is insensitive to receiver height. Therefore, most research works on outdoor radio map estimation~\cite{RadioUNet,FadeNet,Transformerradionet,PPNet} assumed that all receivers are fixed at the same height, and focused on 2D radio map estimation from environmental images depicting the transmitter location and the different kinds of obstacles such as buildings, foliage, etc. For example, in~\cite{RadioUNet}, each pixel of an environmental image has binary value $1$ or $0$ to indicate the existence of an object, while the effect of the object height was not considered. This, however, could lead to inaccurate estimates, as different heights of environmental objects, such as buildings, would have different impacts on signal propagation. The normalized object heights are set as the pixel values of the environmental images in~\cite{FadeNet, Transformerradionet} by normalizing the heights of each type of objects by its $95$ percentile value~\cite{FadeNet} or maximum value~\cite{Transformerradionet}. As a result, the same pixel value could represent different heights for different types of objects, which could lead to inaccurate estimates due to the height ambiguity. Incorporating exact object heights into environmental images is therefore highly desirable for accurate radio map estimation.

For indoor scenarios, existing studies also mainly focused on 2D radio maps~\cite{Pano2RSSI,RadioResUNet,IndoorRSSINet,two-rooms_fixed_transmitter} by assuming all receivers at the same fixed height and all transmitters at another fixed height. However, transmitters, receivers and obstacles usually have different yet comparable heights. As a result, the heights of objects have a significant impact on indoor radio map estimation, as pointed out in~\cite{Pano2RSSI}. Moreover, the authors of~\cite{RadioResUNet} showed that the indoor radio map at a height of $0.2$\,m is significantly different from that at a height of $1$\,m as the layouts seen at different heights are different. Therefore, 3D radio maps that reflect the varying pathloss at different receiver heights should be considered for indoor applications.

To construct 3D radio maps, the 3D propagation environment with object heights is required as the input. The authors of~\cite{2D_to_3D} considered the 3D outdoor layout by slicing it at $5$ discretized heights with a fixed interval as 2D images. Yet, such coarse discretization of object height could lead to large derivations from the actual heights and thus performance loss. This further underscores the importance of accurate height representation. For 3D radio map estimation, existing deep learning based computer vision methods for 2D radio map estimation can be explored. One way is to replace the 2D deep learning operations in existing 2D radio map estimation methods with 3D ones~\cite{2D_to_3D}. Though straightforward, compared to 2D operations, 3D ones add an extra dimension to both kernels and feature maps, which dramatically increases the model size, and hence increases the storage and computational costs. Another way is to treat each 3D radio map as a set of 2D radio maps at a number of discretized heights~\cite{RadioResUNet}. However, estimating the 2D radio maps at different heights requires to train different models, which is computationally costly. Therefore, it is of paramount practical importance to design an efficient 2D deep learning method for 3D radio map estimation.

\begin{figure*}[t]
  \centering
  \includegraphics[width=0.65\linewidth]{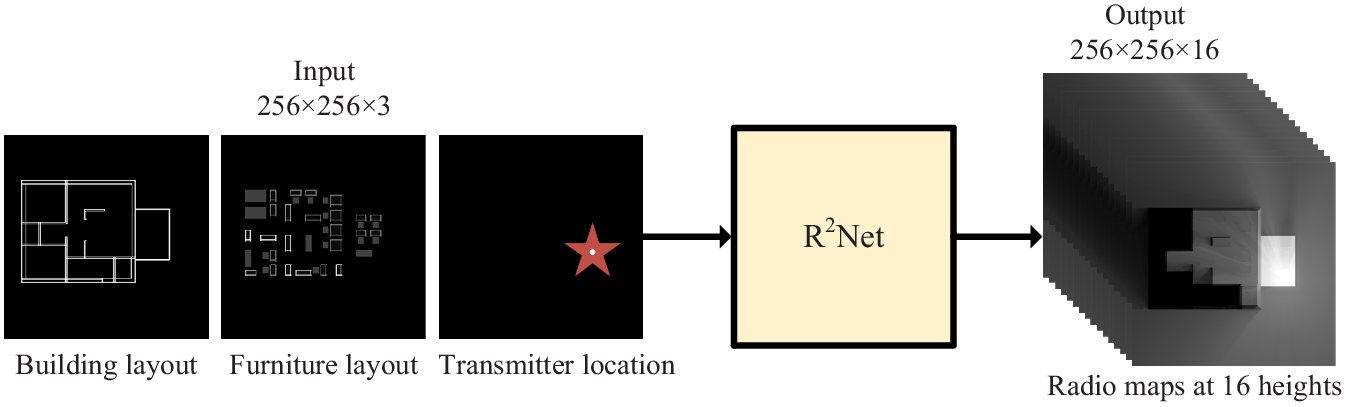}
  \caption{Illustration of the input and the output of the proposed R$^{2}$Net for 3D radio map estimation. Here, the 3D indoor radio map estimation based on our 3DiRM3200 dataset is depicted as an example, where the white dot at the center of the red pentagram highlights the location of the transmitter.}
  \label{3D input and output}
\vspace{-3mm}
\end{figure*}

Moreover, training a deep learning model for 3D radio map estimation requires a complete 3D radio map dataset with pathloss available at every receiver location in each 3D radio map. Regarding 3D radio map datasets, we found two publicly available datasets presented in~\cite{Channel_Charting} and~\cite{Indoor_THz} for indoor scenarios, both of which, however, only contain the measurement results of pathloss at a small number of receiver locations. For outdoor scenarios, a 3D version~\cite{3Dseer} of the widely used 2D radio map dataset, RadioMapSeer~\cite{RadioUNet}, is publicly available, which, however, sets all receivers at the same height of $1.5$\,m.
Therefore, a complete 3D radio map dataset is still lacking.

This paper proposes a novel 2D deep residual learning approach to estimate 3D radio maps by taking into account the impact of object heights. 
The main contributions of this paper are summarized as follows:
\begin{itemize}
\item A 3D indoor radio map dataset (3DiRM3200) consisting of $3,200$ 3D radio maps, $200$ building layout images\footnote{Here, the $200$ building layouts are randomly selected from the publicly available dataset, CubiCasa5K~\cite{CubiCasa5K}, which contains $5,000$ floorplan samples extracted from real estate marketing materials with $36$ distinct building types, including office, hall, library, retail space, bar, basement, bedroom, technical room, etc. As a result, we also focus on these indoor scenarios in this paper.}, $200$ furniture layout images and $16$ transmitter location images for each building is created, to which over $1,000$ labour hours had been dedicated. Apart from 3D radio map estimation, the dataset can also be used for radio map reconstruction, centimeter-level positioning, etc. Our 3DiRM3200 dataset will be available at \url{https://github.com/lighttime2023/3DiRM3200.git}.
\item A 2D height embedding method is proposed to incorporate height information into 2D environmental images via pixel values, which can also distinguish the objects located at a height of $0$m from the non-existence of objects. With this height embedding method, the 3D radio map estimation task can be transformed into a 2D deep learning task by outputting radio maps at different receiver heights through different channels, as illustrated in Fig.~\ref{3D input and output}. As both the input and the output are 2D images, our method requires low computational and storage costs.
\item A novel 2D radio residual network (R$^{2}$Net) is proposed for 3D radio map estimation. As pathloss exhibits different characteristics in indoor and outdoor scenarios,\footnote{It should be noted that for outdoor radio map estimation, we mainly focus on the urban scenarios with transmitters deployed at a similar height with receivers in this paper as the publicly available dataset, RadioMapSeer, only collects radio maps in cities for device-to-device communications.} the R$^{2}$Net is tailored to enhance feature extraction according to pathloss characteristics. For indoor scenarios, R$^{2}$Net-In is proposed by incorporating dropout layers to better capture penetration loss, while for outdoor scenarios, R$^{2}$Net-Out is proposed by adopting atrous spatial pyramid pooling (ASPP) and cascaded residual blocks to extract abundant diffraction loss. In case that only a small training dataset is available, R$^{2}$Net-Outlite is further proposed based on R$^{2}$Net-Out to improve the generalization ability of the model.
\item Extensive quantitative and qualitative experimental results show that the proposed R$^{2}$Net-In outperforms the state-of-the-art benchmarks for 3D indoor radio map estimation in term of estimation accuracy, inference speed, and computational and storage costs. For outdoor radio map estimation, compared to the state-of-the-art benchmarks, R$^{2}$Net-Out achieves the highest estimation accuracy when a large training dataset is available, while R$^{2}$Net-Outlite performs the best in the case with a small training dataset due to its good generalization ability. Experimental results also show that taking into account the height information can improve the accuracy of 2D radio map estimation.
\end{itemize}

The rest of this paper is organized as follows. Section~\ref{section:Related Work} reviews the related works on computer vision based radio map estimation. Section~\ref{section:Preliminaries} presents the preliminaries of key concepts for the ease of understanding this paper. Section~\ref{section:3D Indoor Radio Map Dataset} introduces our created 3D indoor radio map dataset. Section~\ref{R2Net for radio map estimation} presents the proposed R$^{2}$Net for 3D radio map estimation along with its different versions. Section~\ref{Estimating Radio Maps} compares the proposed models and the state-of-the-art benchmarks via experiments. Conclusions are drawn in Section~\ref{Conclusion}.

\section{Related Work}
\label{section:Related Work}
Existing works on exploring computer vision methods for radio map estimation targeted at either outdoor or indoor scenarios. This section reviews outdoor and indoor radio map estimation, respectively, as well as the available datasets.

\vspace{-3mm}
\subsection{Outdoor Radio Map Estimation}
Computer vision based outdoor radio map estimation predicts pathloss based on environmental images~\cite{FadeNet,RadioUNet,Transformerradionet,PPNet,2D_to_3D}. The authors of~\cite{FadeNet,RadioUNet,Transformerradionet,PPNet} studied 2D radio map estimation by assuming the same fixed height for all receivers and employed convolutional neural networks (CNNs), such as U-Net~\cite{UNet} and SegNet~\cite{SegNet}. RadioUNet proposed in~\cite{RadioUNet} estimates 2D outdoor radio maps by cascading two U-Nets. FadeNet~\cite{FadeNet} was developed based on a variant of U-Net by adopting three stride-1 convolutional layers at both the input and the output sides. In~\cite{Transformerradionet}, RadioTrans was proposed by utilizing Transformer-based spread layers in a CNN to model the relationship among the transmitter, the receiver and the environmental objects, while PPNet~\cite{PPNet} employs SegNet~\cite{SegNet} instead. In addition to the above works on 2D outdoor radio map estimation, the authors of~\cite{2D_to_3D} extended RadioUNet~\cite{RadioUNet} to 3D outdoor radio map estimation by replacing 2D deep learning operations with 3D ones at the price of significantly increased computational and storage costs due to the greatly increased number of model parameters to be trained. Despite the successes shown by computer vision methods, how to enhance feature exaction based on signal propagation characteristics is underexplored. This paper will handle this crucial point by designing models according to pathloss characteristics.

For the input of the existing methods, the environmental images at a fixed height were used in~\cite{RadioUNet}, by simply setting the value of the pixel at which a building or a transmitter is located as $1$ and the others as $0$s. The heights of objects were considered in~\cite{FadeNet,Transformerradionet,PPNet} and represented by pixel values. To equalize the height information for model convergence, FadeNet~\cite{FadeNet} and RadioTrans~\cite{Transformerradionet} normalize the object heights of each type of material by its $95$ percentile value and maximum value, respectively. This leads to the same pixel value representing different heights for the objects with different materials, i.e., inaccurate input of object heights. In addition, the objects located at a height of $0$\,m cannot be distinguished from the non-existence of the object. To address these challenges, this paper will propose a method to embed height information in 2D environmental images.

\vspace{-3mm}
\subsection{Indoor Radio Map Estimation}
Similarly, previous works on indoor radio map estimation also mainly focused on 2D radio map estimation by assuming that all the receivers are at the same height~\cite{Pano2RSSI,IndoorRSSINet,RadioResUNet,two-rooms_fixed_transmitter}. RSSI-Net~\cite{Pano2RSSI} adopts an optimized variant of U-Net~\cite{UNet} with four convolutional layers in each downsampling/upsampling layer to improve the estimation accuracy, while the canonical U-Net uses two convolutional layers. By cascading two RSSI-Nets, IndoorRSSINet was developed in~\cite{IndoorRSSINet}. RadioResUNet~\cite{RadioResUNet} was proposed based on RadioUNet~\cite{RadioUNet}, which adopts ResNet~\cite{resnet} in the encoder to address the model degradation issue caused by the vanishing gradient problem in deep learning. The authors of~\cite{two-rooms_fixed_transmitter} aimed to reduce the number of required training samples in similar environments, and proposed a deep transfer learning method based on CNN. Similar to the outdoor cases, the computer vision models adopted in the aforementioned works barely take the signal propagation characteristics into consideration, leaving room for improvement. This paper will propose specific models for outdoor and indoor scenarios, respectively.

It should be noted that, in contrast to outdoor scenarios, the indoor radio maps observed by the receivers at different heights could be significantly different~\cite{RadioResUNet}. Estimating 3D radio map is, therefore, essential for indoor cases. For 3D radio map estimation, the existing computer vision methods for 2D radio map estimation can be employed straightforwardly by treating a 3D radio map as a set of 2D radio maps at a number of discretized heights~\cite{RadioResUNet} and training different models for estimating the radio maps at different heights. Another way is to replace the 2D deep learning operations in the existing 2D radio map estimation methods with 3D ones~\cite{2D_to_3D}. Both methods significantly increase the amount of parameters, incurring huge computational and storage costs of model training. By embedding height information into 2D images, this paper will propose a 2D deep residual learning approach for 3D radio map estimation at low cost.

\vspace{-3mm}
\subsection{Radio Map Dataset}
To train deep learning based computer vision methods to estimate radio maps, a complete radio map dataset with pathloss value available at every possible receiver location is indispensable. RadioMapSeer~\cite{RadioUNet} is a publicly available dataset with $56,000$ simulated 2D outdoor radio maps, which has been widely used for 2D radio map estimation and reconstruction. For indoor scenarios, there are two publicly available datasets~\cite{Channel_Charting,Indoor_THz}, both of which only contain a small amount of measured pathloss at some receiver locations, leading to a lot of blank spaces in the radio maps. Specifically, the dataset in~\cite{Channel_Charting} provides the measured pathloss along some meandering trajectory for $11$ practical scenarios, and the dataset in~\cite{Indoor_THz} provides the pathloss at $15$ receiver locations in a hallway. Recently, the authors of~\cite{WiSegRT-3ddata} attempted to create a complete 3D indoor radio map dataset by simulating the pathloss at $15\times15\times15$ receiver locations in $10$ scenes. However, only the environmental data are open and simulations need to be run to obtain radio maps. A complete 3D radio map dataset with sufficient samples for training computer vision methods is still lacking. As indoor radio maps are more sensitive to height information, this paper will create a complete 3D indoor radio map dataset, which contains $3,200$ samples in $200$ scenes, each with $16\times256\times256$ receiver locations.

\section{Preliminaries}
\label{section:Preliminaries}
This section presents the prerequisites for understanding the paper, including dominant path model (DPM) for creating our 3DiRM3200 dataset, U-Net as the foundational model of our proposed R$^{2}$Net, ResNet for extracting diffraction loss and atrous spatial pyramid pooling (ASPP) for extracting multi-scale features to handle the obstacles of diverse sizes.

\vspace{-3mm}
\subsection{DPM}
\label{subsection:DPM}
DPM~\cite{DPM} is a classical physical simulation method, which determines the dominant path between the transmitter and the receiver and simulates the pathloss along the path. In complex indoor environments where multiple interactions occur and the number of interactions is usually large and random, DPM is able to capture all the interactions along the dominant path and hence can obtain the pathloss accurately. Specifically, DPM obtains pathloss $P_{\textup{L}}$ by
\begin{equation}
\left(P_{\textup{L}}\right)_{\rm dB}{=}20\log\left(\frac{4\pi}{\lambda}\right)+10p\log\left(l\right)+\sum_{i=1}^{N_f} f\left(\varphi_i,i\right)+\sum_{j=1}^{N_t} t_j-\Omega,
\end{equation}
where $\lambda$ is the wavelength, $p$ is the pathloss exponent, and $l$ is the path length. $f(\varphi_i,i)$ is the loss incurred by the $i$th interaction that changes the signal propagation direction. $\varphi_i$ is the corresponding propagation angle difference resulting from the $i$th interaction, and $N_f$ is the number of interactions. $t_j$ is the transmission loss caused by the $j$th wall, and $N_t$ is the number of walls. $\Omega$ is the waveguiding factor.

\vspace{-3mm}
\subsection{U-Net}
\begin{figure}[t]
  \centering\includegraphics[width=0.85\linewidth]{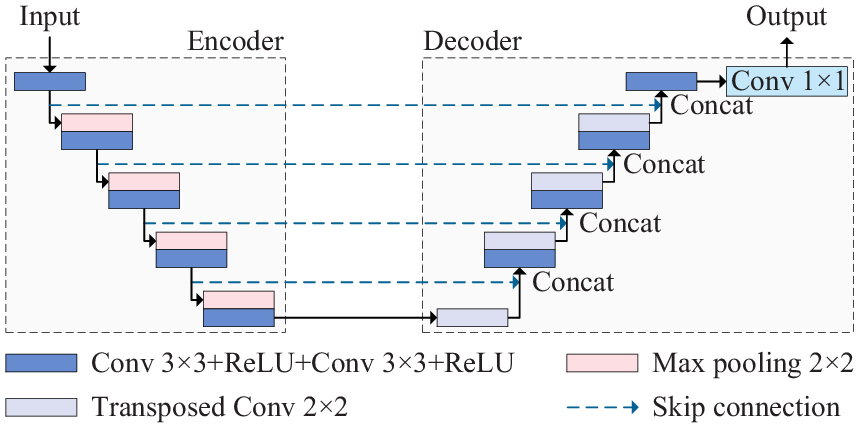}
  \caption{Illustration of a typical U-Net that serves as the foundational model of our proposed R$^{2}$Net.}
  \label{U-Net architecture}
  \vspace{-3mm}
\end{figure}
\begin{figure}[t]
  \centering
  \includegraphics[width=0.65\linewidth]{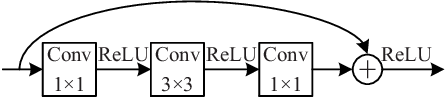}
  \caption{{Illustration of a typical residual block that inspires our design for extracting diffraction loss features.}}
  \label{A residual block}
    \vspace{-3mm}
\end{figure}

U-Net~\cite{UNet} has been widely adopted in radio map estimation and showed superior performance~\cite{FadeNet,RadioUNet,Pano2RSSI,IndoorRSSINet,RadioResUNet}, which consists of an encoder and a decoder. A typical U-Net model is shown in Fig.~\ref{U-Net architecture}.

\subsubsection{Encoder}
The encoder extracts features through a series of convolutional layers, activation functions and pooling layers, gradually reducing the resolution of the input. In a convolutional layer, an input feature map is convolved with a kernel and a bias is added. An activation function is applied to introduce nonlinearity, for which ReLU is chosen typically in U-Net. A pooling layer downsamples feature maps. In a U-Net, max pooling~\cite{maxpooling} is commonly used.

\subsubsection{Decoder}
The decoder employs transposed convolutional layers, convolutional layers and activation functions to increase the resolution of the feature map.
Specifically, the canonical U-Net adopts a transposed convolutional layer with stride $2$ in the decoder.

\subsubsection{Skip Connection}
Skip connections, as represented by the blue arrows in Fig.~\ref{U-Net architecture}, concatenate the cropped feature maps from the encoder with the corresponding ones at the decoder, which can address the gradient vanishing problem and enable the training of deep networks. The cropping process is necessary because the resolution of a feature map at the encoder could be higher than that at the decoder due to the loss of border pixels after convolution when the size of the input feature map is not divisible by the kernel size.

\vspace{-3mm}
\subsection{ResNet}
ResNet~\cite{resnet} introduces a residual block to ease the training of very deep networks with affordable training time. A typical residual block is illustrated in Fig.~\ref{A residual block}, which stacks the $1{\times}1$, $3{\times}3$ and $1{\times}1$ convolutional layers. Two $1{\times}1$ convolutional layers are responsible for reducing and increasing the number of channels, respectively, leaving the $3{\times}3$ convolutional layer with a smaller number of channels to reduce training time. In addition, $1{\times}1$ convolutional layers and ReLU introduce nonlinearity, helping extract nonlinear features. Moreover, the residual connection that directly connects the input of a certain layer and the output of a later layer provides an alternative path for the gradient flow, which resolves the gradient vanishing problem in training substantially deep networks.

\begin{figure}[t]
  \centering
  \includegraphics[width=0.95\linewidth]{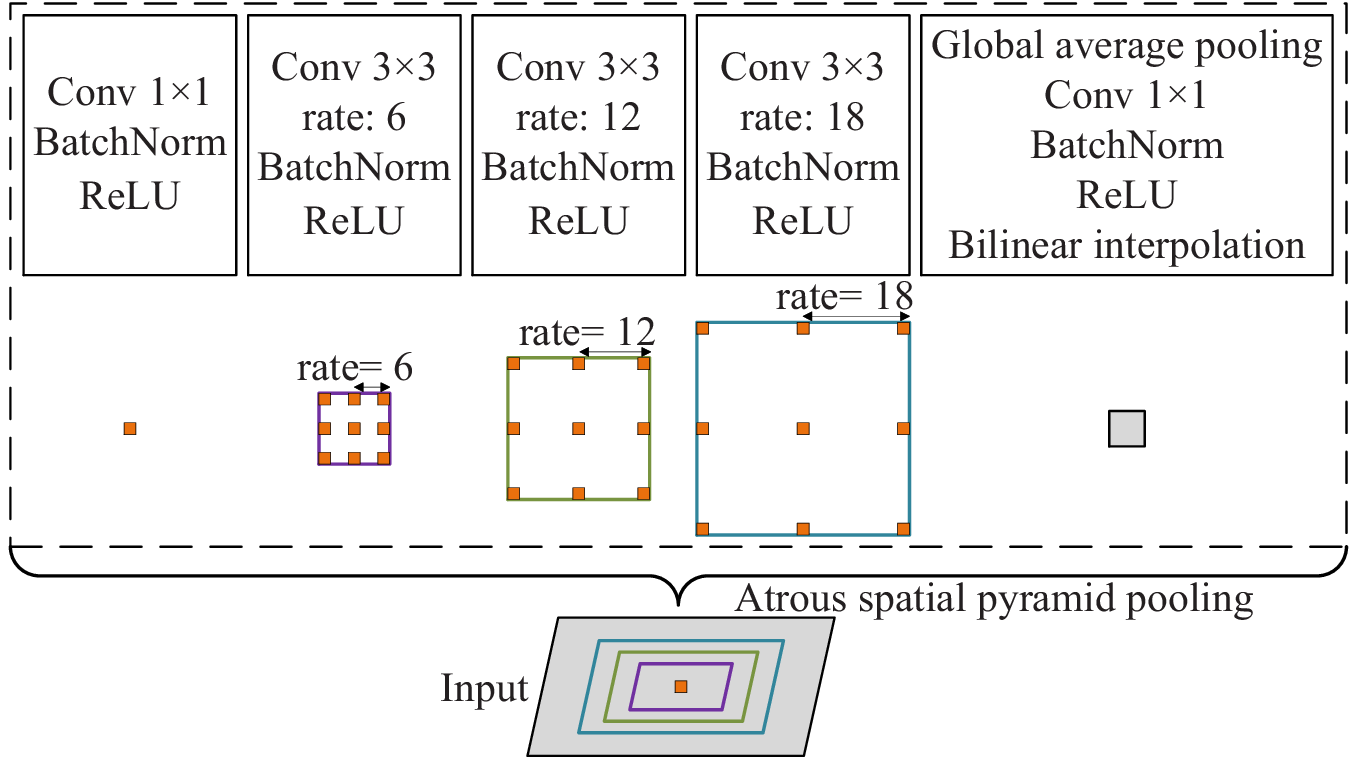}
  \caption{Illustration of ASPP.}
  \label{ASPP}
  \vspace{-3mm}
\end{figure}

\vspace{-3mm}
\subsection{ASPP}
ASPP~\cite{DeepLab-ASPP} is designed to handle the objects of different sizes by employing parallel atrous convolutions at multiple dilation rates to capture multi-scale information. As illustrated in Fig.~\ref{ASPP}, ASPP consists of a $1\times1$ convolutional layer and three $3\times3$ convolutional layers with atrous rates $6$, $12$ and $18$, respectively, for extracting multi-scale local features. To capture global features, global average pooling is adopted to take the average of the feature maps.

\section{3D Indoor Radio Map Dataset}
\label{section:3D Indoor Radio Map Dataset}

\begin{figure*}[t]
\centering
\subfloat[Building layout]{\includegraphics[width=0.14\textwidth]{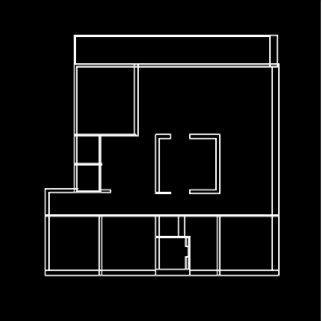}\label{Building layout}}
\subfloat[Furniture layout]{\includegraphics[width=0.14\textwidth]{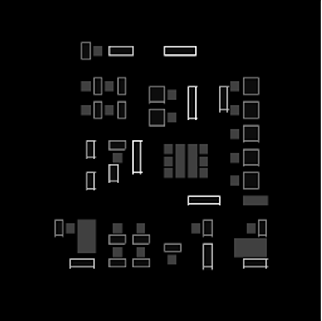}\label{Furniture layout}}
\subfloat[Location of Tx 1]{\includegraphics[width=0.14\textwidth]{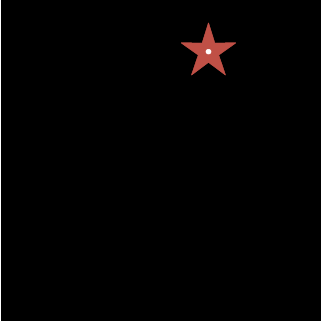}\label{Transmitter location}}
\subfloat[Radio map at a height of $0.5$\,m (Tx 1)]{\includegraphics[width=0.14\textwidth]{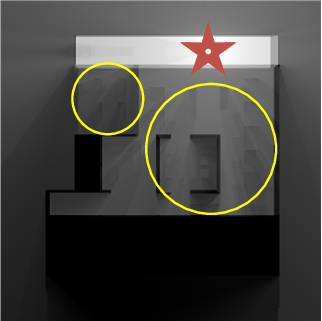}\label{Radio map05}}
\subfloat[Radio map at a height of $1.5$\,m (Tx 1)]{\includegraphics[width=0.14\textwidth]{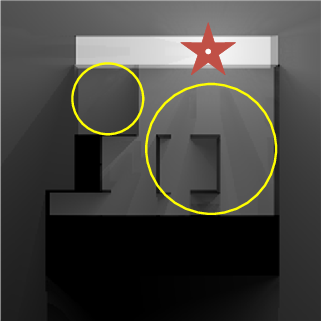}\label{Radio map15}}
\subfloat[Location of Tx 2]{\includegraphics[width=0.14\textwidth]{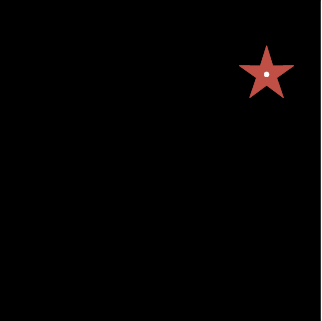}\label{Transmitter location2}}
\subfloat[Radio map at a height of $0.5$\,m (Tx 2)]{\includegraphics[width=0.14\textwidth]{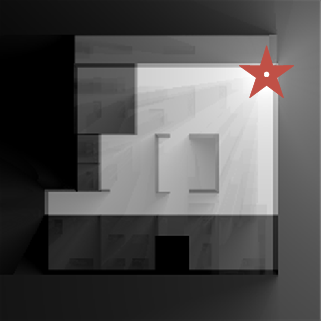}\label{Radio map2_05}}
\caption{Images from sample No. 120 of 3DiRM3200, including (a) building layout depicting the locations of walls, (b) furniture layout presenting the locations and shapes of furniture, two exemplary transmitter (Tx) locations (c) Tx 1 and (f) Tx 2, exemplary radio maps at a height of (d) $0.5$\,m and (e) $1.5$\,m with Tx 1, and at a height of (g) $0.5$\,m with Tx 2. The white dot at the center of the red pentagram highlights the transmitter location. The yellow circles highlight the impact of furniture on the radio maps at different heights.}
\label{Dataset}
\vspace{-3mm}
\end{figure*}

This section introduces our created 3D indoor radio map dataset (3DiRM3200), which consists of $3,200$ 3D radio maps, $200$ images of building layouts, $200$ images of furniture layouts and $3,200$ images depicting $3,200$ randomly generated transmitter locations. Specifically, for each building, a set of $16$ 3D radio maps is created, each corresponding to one transmitter location, and one 3D radio map is a collection of $16$ 2D radio maps at $16$ different receiver heights. Let us take sample No. 120 in our 3D radio map dataset for illustration. Figs.~\ref{Dataset}(a), (b), (c) and (f) showcase its building layout, furniture layout, the first randomly generated transmitter location (Tx 1) and the second randomly generated transmitter location (Tx 2), respectively, where the height information is embedded in the images as pixel values. Figs.~\ref{Dataset}(d) and (e) illustrate the radio maps at a height of $0.5$\,m and $1.5$\,m with Tx 1, respectively, and Fig.~\ref{Dataset}(g) showcases the radio map at a height of $0.5$\,m with Tx 2. Note that even for the same building with the same furniture layout and the same receiver height, a significant difference can be observed between the radio map with Tx 1 shown in Fig.~\ref{Dataset}(d) and the radio map with Tx 2 shown in Fig.~\ref{Dataset}(g) due to the effect of transmitter location.

\vspace{-2mm}
\subsection{Environmental Images}
\label{subsection Environmental Images}
Environmental images depict the radio propagation environment, which are the input of radio map estimation. The environmental images in our dataset include the images showing the building layout, the furniture layout and the transmitter location. As illustrated in Fig.~\ref{Dataset}(a), a building layout depicts the locations of walls. For each building, furniture is drawn manually with random locations and sizes, and the transmitter is randomly located in the building. In contrast to 2D radio map estimation that ignores the impact of the heights of obstacles and transmitters/receivers, and thus could degrade the accuracy of radio map estimation, in our dataset, the heights of walls, furniture and the transmitter are embedded into 2D environmental images by transforming them into pixel values. All environmental images have the same size of $20.44$\,m$\times$$20.44$\,m with $256\times256$ pixels and a pixel interval of $8$\,cm. The details of creating the dataset are presented below.

\subsubsection{Environmental Data}
To create environmental images, the environmental data, e.g., the locations and heights of the objects in a propagation environment are required. To this end, we randomly select $200$ floorplan samples from the publicly available dataset, CubiCasa5K~\cite{CubiCasa5K}, which consists of $5,000$ floorplans from real estate marketing materials. However, the materials of the objects in the floorplans, such as walls, doors, windows and furniture, are not provided. As the materials of objects largely affect the pathloss and hence the radio map, we keep the walls only in each floorplan and manually draw the furniture. Specifically, we first convert the floorplan from its original SVG format to an AutoCAD file in DXF format by using the SVG editor Inkscape~\cite{Inkscape}. The DXF file is then imported to the WinProp simulation software~\cite{WinProp} to obtain the environmental data file with walls only. The height of all walls is set at $3$\,m, and the material property of walls is set as brick. Furniture is then drawn in WinProp manually in a random shape with the height randomly set at either $0.5$\,m, $1$\,m, $1.5$\,m, or $2$\,m. The material of furniture is fir wood. The height of the transmitter takes values from $0$\,m to $3$\,m with an interval of $0.2$\,m. That is, there are $16$ transmitter locations in total for each building. The environmental data are then exported from WinProp.

\subsubsection{Height Embedding}
\label{heightembedding}
To embed height information into 2D environmental images, we represent the height by the normalized pixel value between $0$ and $1$, and propose a new normalization method. Particularly, a pixel where nothing is located has value $0$. To distinguish the objects located at a height of $0$\,m from the non-existence of objects, we propose to set the value of a pixel where a wall, a piece of furniture or a transmitter is located as
\begin{equation}\label{eq:heightembedding}
v_{env}=(h_{env}+\epsilon)/(h_{env,max}+\epsilon),
\end{equation}
where $h_{env,max}$ denotes the maximum height of objects in the environment and $\epsilon\ll h_{env,max}$ is the added bias. In our case, the maximum height $h_{env,max}$ is the same as the height of walls, i.e., $h_{env,max}=3$\,m, and the bias is chosen as $\epsilon=0.1$\,m, i.e., the height resolution of the 3D radio map. According to the above 2D height embedding method, the environmental images depicting the building layout, the furniture layout and the transmitter location can be obtained from WinProp based on the environmental data.

\vspace{-2mm}
\subsection{3D Radio Maps}
\label{subsection:3D Indoor Radio Map}
The 3D radio maps in our dataset are obtained using the WinProp simulation software~\cite{WinProp}, where DPM is chosen to obtain pathloss because it can trace complex signal interactions in complex indoor environments. A 3D radio map is a set of $16$ 2D radio maps at $16$ receiver heights varying from $0.5$\,m to $2$\,m with an interval of $0.1$\,m. This is based on the practical consideration that a receiver usually has the height between $0.5$\,m and $2$\,m. For a 2D radio map at some height, as shown in Figs.~\ref{Dataset}(d), (e) and (g), it has the fixed size of $20.44$\,m$\times$$20.44$\,m with $256\times256$ pixels. That is, there are $16\times256\times256$ receiver positions in total per each 3D radio map. The pixel interval length of each 2D radio map is set as $8$\,cm so that it can be potentially used for high-precision indoor localization. The process of generating 3D radio maps is detailed below.

\subsubsection{Parameter Settings}
As a number of wireless communication systems operate at $5.9$\,GHz, such as the WiFi system adopting IEEE 802.11n, 802.11ac and/or 802.11ax standards, a signal bandwidth $W=10$\,MHz in the $5.9$\,GHz band is set.

The transmit power is fixed at $P_{\textup{Tx}} = 23$\,dBm. The power spectral density of thermal noise is $N_0 =-174$\,dBm/Hz, and an idealistic noise figure at the receiver side is assumed as $N_{\textup{fig}}=0$\,dB~\cite{RadioUNet}.
The noise floor $N_{\textup{flr}}$ is then obtained as
\begin{equation}
(N_{\textup{flr}})_{\rm dB} =  10\log_{10} W N_0 + (N_{\textup{fig}})_{\rm dB}.
\end{equation}
Since it is usually difficult to decode a wireless signal if the received signal strength (RSS), given by
\begin{equation}
(P_{\textup{Rx}})_{\rm dB} = (P_{\textup{Tx}})_{\rm dB} - (P_{\textup{L}})_{\rm dB},
\end{equation}
is below the noise floor $N_{\textup{flr}}$, we are only interested in the pathloss below some value such that the RSS $P_{\textup{Rx}}$ is no less than $N_{\textup{flr}}$, i.e., $-(P_{\textup{L}})_{\rm dB} \geq (N_{\textup{flr}})_{\rm dB} - (P_{\textup{Tx}})_{\rm dB}$. As the pathloss output by the WinProp simulation software is $-P_{\textup{L}}$, we define
\begin{equation}
    (L)_{\rm dB} \triangleq - (P_{\textup{L}})_{\rm dB},
\end{equation}
and refer to $L$ as pathloss hereafter.
According to above discussion, the pathloss threshold $L_{\textup{thr}}$ can be set as
\begin{equation}
(L_{\textup{thr}})_{\rm dB} = (N_{\textup{flr}})_{\rm dB} - (P_{\textup{Tx}})_{\rm dB},
\end{equation}
which is $-127$\,dB according to the above parameter settings.

Considering that low-power communication systems, such as satellite communications and wireless sensor networks, usually suffer from severe power attenuation, attention should be paid to the RSS both above and a bit below the noise floor. Similar to~\cite{RadioUNet}, we truncate the pathloss below another smaller threshold $L_{\textup{trnc}}$ than $L_{\textup{thr}}$ in our 3DiRM3200 dataset, which is referred to as analytic noise floor and chosen as
\begin{equation}
(L_{\textup{trnc}})_{\rm dB} \triangleq (5(L_{\textup{thr}})_{\rm dB}-(M_{\textup{1}})_{\rm dB})/4,
\end{equation}
where $M_{\textup{1}}$ is the maximum pathloss in the dataset~\cite{RadioUNet}.

\subsubsection{Pathloss Conversion}
To represent pathloss in a radio map in the form of an image, the pathloss in dB scale needs to be converted to pixel values. By ignoring the pathloss below $L_{\textup{trnc}}$, the ground-truth pathloss $L$ is normalized as a pixel value between $0$ and $1$ by
\begin{equation}\label{eq:pathloss_normalization}
v_{\textup{L}}=\max\left\{\frac{(L)_{\rm dB}-(L_{\textup{trnc}})_{\rm dB}}{(M_{\textup{1}})_{\rm dB}-(L_{\textup{trnc}})_{\rm dB}},0\right\}.
\end{equation}
In particular, $v_{\textup{L}}=0$ represents that the pathloss is below or equal to the analytic noise floor $L_{\textup{trnc}}$, and $v_{\textup{L}}=1$ indicates the maximum pathloss $M_{\textup{1}}$. Such normalization aligns well with the common practice of deep learning based computer vision methods that they output probabilities between $0$ and $1$.

\section{R$^{2}$Net for Radio Map Estimation}
\label{R2Net for radio map estimation}
A 2D deep residual learning method, radio residual network (R$^{2}$Net), is proposed in this section to estimate 3D radio maps. As the diffraction loss caused by furniture or buildings is an important characteristic for both indoor and outdoor scenarios, the proposed R$^{2}$Net enhances the feature extraction for diffraction loss. Furthermore, to precisely extract features according to indoor or outdoor pathloss characteristics, the proposed R$^{2}$Net is tailored in different scenarios. In indoor scenarios, the penetration loss caused by walls leads to severe signal attenuation and dominates the pathloss, while in outdoor scenarios, the diffraction loss largely affects the pathloss as massive buildings bring sudden changes to signal attenuation and propagation direction. Since the pathloss characteristics are different in indoor and outdoor scenarios, R$^{2}$Net-In is proposed for indoor radio map estimation, and R$^{2}$Net-Out is proposed for outdoor radio map estimation. A light variant, R$^{2}$Net-Outlite, is also proposed to accommodate the case with a small training dataset. Before presenting the proposed R$^{2}$Net in detail, let us first illustrate how we transform the 3D radio map estimation task into a 2D task.

\begin{figure*}[t]
  \centering
  \includegraphics[width=0.8\linewidth]{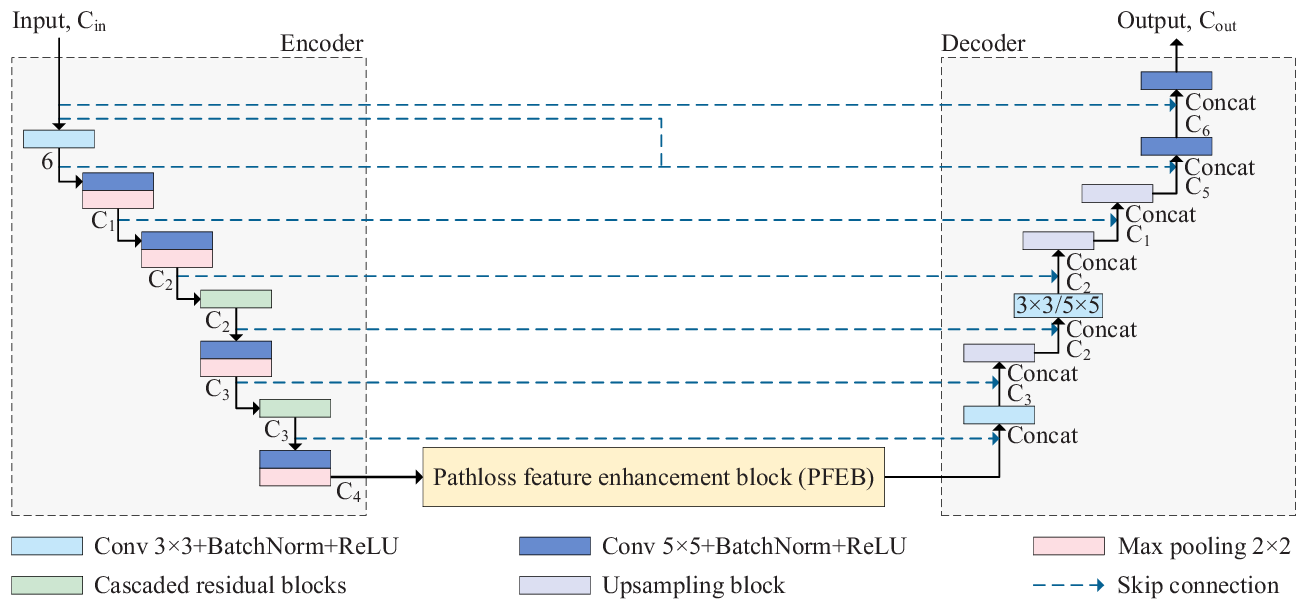}
  \caption{R$^{2}$Net architecture.}
  \label{R2Net}
  \vspace{-2mm}
\end{figure*}

\begin{figure*}[t]
  \centering
  \includegraphics[width=1\linewidth]{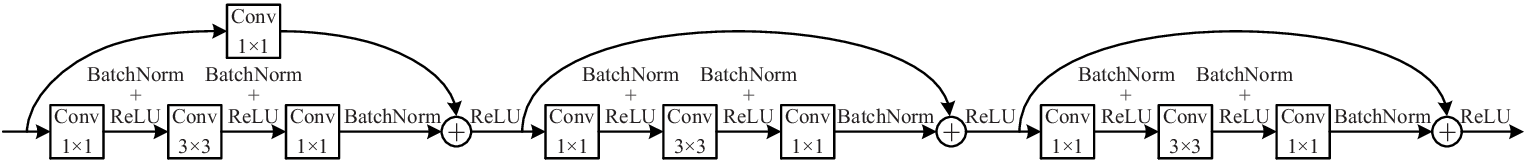}
  \caption{Cascaded residual blocks.}
  \label{A bottleneck residual block}
    \vspace{-3mm}
\end{figure*}

\vspace{-3mm}
\subsection{3D-to-2D Task Transform}
\label{Modeling of 3D Indoor Radio Map Estimation}
In order to enable efficient 3D radio map estimation with a small model, we transform the 3D estimation task into a task that can be handled by 2D computer vision methods, whose input and output are both 2D images. Specifically, by adopting the height embedding method proposed in Section~\ref{subsection Environmental Images}, which represents the heights of environmental objects by the corresponding pixel values, the height information is embedded into 2D images and these 2D images can serve as the input. For the output, to represent 3D radio map by 2D images, the 3D radio map is discretized into a set of 2D radio maps at different heights, and each 2D radio map corresponds to one output channel, as shown in Fig.~\ref{3D input and output}. The 3D radio map estimation task is now transformed into a 2D computer vision task.

\vspace{-3mm}
\subsection{R$^{2}$Net Architecture}
As illustrated in Fig.~\ref{R2Net}, R$^{2}$Net is designed based on U-Net that consists of an encoder and a decoder, in addition to which, a pathloss feature enhancement block (PFEB) between the encoder and the decoder is proposed to enhance pathloss feature extraction.

\subsubsection{Encoder}
The encoder extracts the pathloss features through a series of convolutional layers, batch normalization, ReLU, cascaded residual blocks and max pooling. In both indoor and outdoor scenarios, the diffraction loss caused by furniture or buildings is an important characteristic that needs to be carefully considered~\cite{RadioUNet,FadeNet,diffraction,2D_to_3D,diffraction2,diffraction3}. Particularly, the diffraction loss causes sudden attenuation of signal power and change of propagation direction, leading to nonlinear impact on the pathloss. To extract nonlinear features, in the encoder of R$^{2}$Net as shown in Fig.~\ref{R2Net}, three cascaded residual blocks presented in Fig.~\ref{A bottleneck residual block} are adopted to replace the convolutional layer in the encoder of the standard U-Net. Particularly, ReLU and $1\times1$ convolutional layers in the adopted cascaded residual blocks introduce nonlinearity for better extracting nonlinear features. The residual blocks adopted here are also able to handle the gradient vanishing problem.

\subsubsection{PFEB}
As the pathloss characteristics are different in indoor and outdoor scenarios, we propose different PFEBs to extract the corresponding pathloss features. For indoor radio map estimation, PFEB-In is proposed to enhance the capture of penetration loss, while for outdoor radio map estimation, PFEB-Out is designed to extract abundant diffraction loss. The details will be presented in Sections V-C and V-D.

\subsubsection{Decoder}
The decoder employs upsampling blocks, convolutional layers, batch normalization and ReLU to predict the radio map. Since PFEB-In and PFEB-Out are customized for indoor and outdoor scenarios, respectively, different upsampling blocks are required accordingly to restore the resolution in the decoder of R$^{2}$Net-In and R$^{2}$Net-Out, which will be detailed later in Section~\ref{sec-R2Net-In} and Section~\ref{sec-R2Net-Out}.

\subsubsection{Skip Connection}
Similar to the standard U-Net, skip connections are adopted in R$^{2}$Net, connecting the encoder and the decoder. As represented by the blue arrows in Fig.~\ref{R2Net}, skip connections copy the feature maps from the encoder and concatenate them to the corresponding feature maps at the decoder. By carefully designing the convolutional layers in the encoder and the decoder of R$^{2}$Net, the number of pixels in a feature map at the encoder is ensured to be the same as that at the decoder, as a result of which, cropping used in the conventional U-Net is not needed in our R$^{2}$Net.

\subsubsection{Numbers of Channels}
As the effect of an environmental object on the pathloss depends on its material, to distinguish different materials of objects, the number of input channels $C_{\textup{in}}$ is equal to the number of types of object materials in the dataset plus $1$, with the extra one channel dedicated to the transmitter location image. The number of output channels $C_{\textup{out}}$ is the same as the number of 2D radio maps collected in a 3D radio map, i.e., the number of receiver heights. It is obvious that 2D radio map estimation is a special application of our 3D radio map estimation network with $C_{\textup{out}}=1$. Particularly, if 3D radio map estimation is conducted based on our 3DiRM3200 dataset, we have $C_{\textup{in}}=3$ and $C_{\textup{out}}=16$. $C_{\textup{1}}$, $C_{\textup{2}}$, $C_{\textup{3}}$, $C_{\textup{4}}$, $C_{\textup{5}}$ and $C_{\textup{6}}$ in Fig.~\ref{R2Net} are the numbers of output channels in the corresponding layers.

\subsubsection{Loss Function}
The loss function adopted to train our R$^{2}$Net is the mean square error (MSE) of the radio map estimates, given by
\vspace{-2mm}
\begin{equation}
    \textup{Loss}={\frac{1}{N_{\textup{train}}}}\sum_{i=1}^{N_{\textup{train}}}\lVert\widetilde{\mathbf{V}}(i)-{\mathbf{V}}(i)\rVert^2_2,
    \label{eq:MSE}
\end{equation}
where $\widetilde{\mathbf{V}}(i)$ is the estimate of the $i$-th radio map, and ${\mathbf{V}}(i)$ is the ground truth. $N_{\textup{train}}$ is the number of training samples.

\subsection{R$^{2}$Net-In for Indoor Scenarios}
\label{sec-R2Net-In}
For indoor radio map estimation, R$^{2}$Net-In is proposed, in which PFEB-In is designed to enhance the extraction of indoor pathloss, such as penetration loss, and all upsampling blocks adopt nearest-neighbor interpolation to mitigate the impact of the information loss brought by PFEB-In. The details are presented in the following.

\begin{figure*}[t]
  \centering
  \includegraphics[width=0.9\linewidth]{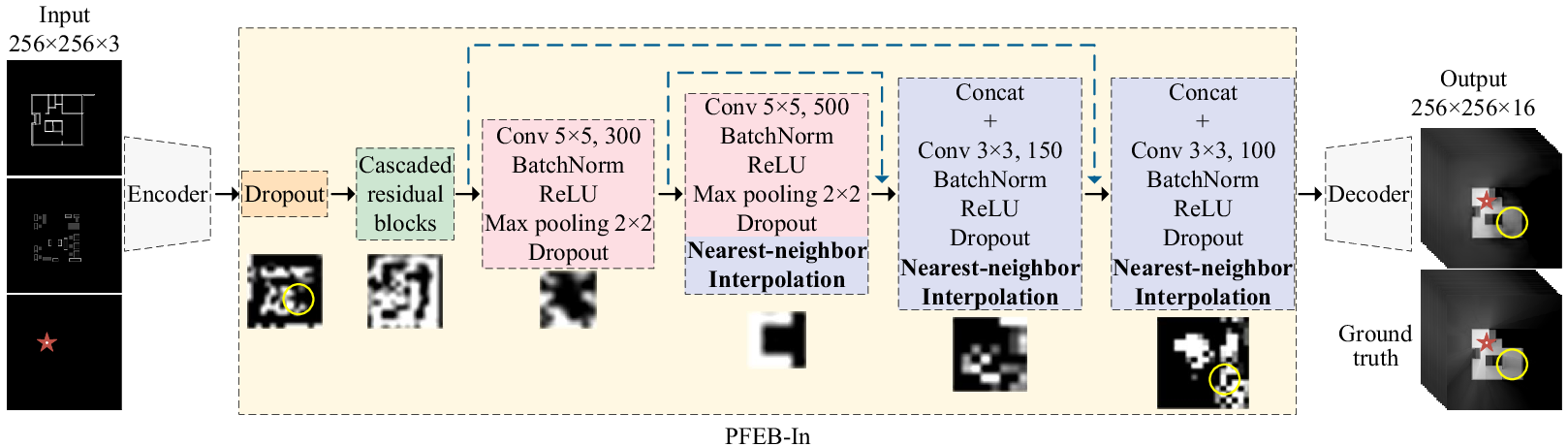}
  \caption{R$^{2}$Net-In architecture.}
  \label{R2Net-In}
  \vspace{-4mm}
\end{figure*}

\subsubsection{PFEB-In}
In indoor scenarios, penetration loss makes a significant contribution to the pathloss~\cite{penetration2,penetration}. As a result, the 3D indoor radio map varies significantly with the transmitter location, even for the same building with the same furniture layout, as can be observed from Figs.~\ref{Dataset}(d) and (g). Such a phenomenon indicates that a good model should have strong generalization ability. PFEB-In is proposed based on this observation, which adopts dropout layers with a dropout probability of $0.3$ on $16\times16$, $8\times8$ and $4\times4$ feature maps by stochastically ``dropping out'' neurons~\cite{dropout} to avoid overfitting, as illustrated in Fig.~\ref{R2Net-In}. 

As the adopted dropout layers and max pooling layers lead to information loss, nearest-neighbor interpolation is adopted afterwards to assign the value of the nearest pixel to the corresponding pixel in the output feature map~\cite{deconvolution_artifacts}. As shown in Fig.~\ref{R2Net-In}, compared to the first feature map of PFEB-In, the output feature map of PFEB-In is more similar to the ground truth. For instance, the area highlighted by the yellow circle in the first feature map is mostly black, while containing more white pixels in the output feature map, i.e., more similar to the ground truth. As can be seen in the figure, this area is far away from the transmitter and there are many walls between it and the transmitter. Yet the wireless signal can still reach this area due to a large number of diffractions. That is, signal propagation is complex in this case, and consequently, the effectiveness of our PFEB-In corroborates its ability to extract complex indoor pathloss features.

\begin{figure}[t]
\centering
\subfloat[R$^{2}$Net-In]{\includegraphics[width=0.33\linewidth]{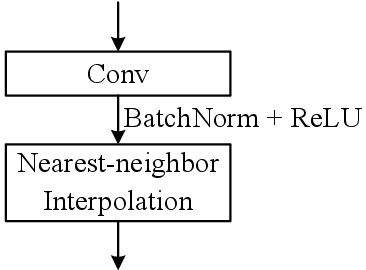}\label{ConvNNI}}
\hspace{1mm}
\subfloat[R$^{2}$Net-Out]{\includegraphics[width=0.205\linewidth]{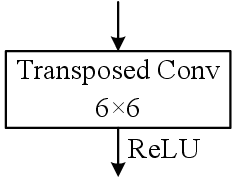}\label{TConv}}
\caption{Upsampling blocks for (a) R$^{2}$Net-In and (b) R$^{2}$Net-Out.}
\label{interpolation}
  \vspace{-3mm}
\end{figure}

Similarly, nearest-neighbor interpolation is adopted for all upsampling blocks in the decoder of R$^{2}$Net-In. Specifically, an upsampling block consists of convolutional layers, batch normalization, ReLU and nearest-neighbor interpolation, as illustrated in Fig.~\ref{interpolation}(a). To capture global features, feature maps with higher resolution employ convolutional layers with larger kernel size. Specifically, in the decoder, the first upsampling block uses a $3\times3$ convolutional layer, and the other two upsampling blocks close to the output adopt a $5\times5$ convolutional layer. Moreover, after the first upsampling block in the decoder, as depicted in Fig.~\ref{R2Net}, a $3\times3$ convolutional layer is employed to capture local features.

\subsubsection{Numbers of Channels}
To maintain a relatively small model, the numbers of channels in R$^{2}$Net-In are set as $C_{\textup{1}}=40$, $C_{\textup{2}}=60$, $C_{\textup{3}}=100$, $C_{\textup{4}}=150$, $C_{\textup{5}}=20$ and $C_{\textup{6}}=20$.

\vspace{-3mm}
\subsection{R$^{2}$Net-Out for Outdoor Scenarios}
\label{sec-R2Net-Out}
Based on the proposed R$^{2}$Net architecture, R$^{2}$Net-Out is designed to estimate outdoor radio maps, in which PFEB-Out concentrates on extracting abundant features of diffraction loss that dominants the outdoor pathloss. The detailed architecture is described below.

\begin{figure*}[t]
  \centering
  \includegraphics[width=0.9\linewidth]{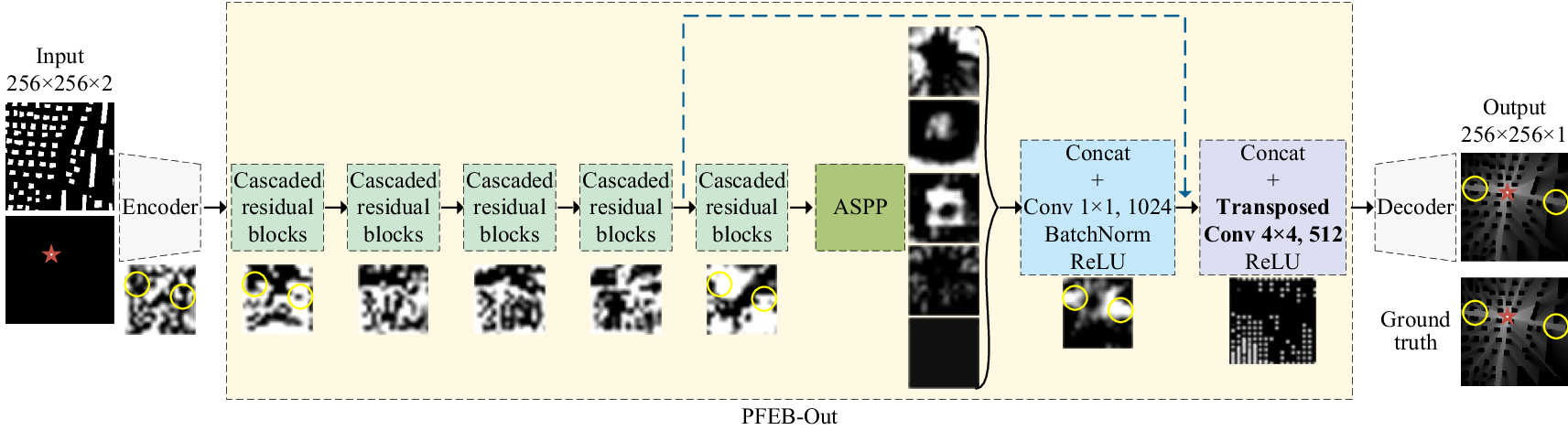}
  \caption{R$^{2}$Net-Out architecture.}
  \label{R2Net-Out}
\end{figure*}

\begin{figure*}[t]
  \centering
  \includegraphics[width=0.9\linewidth]{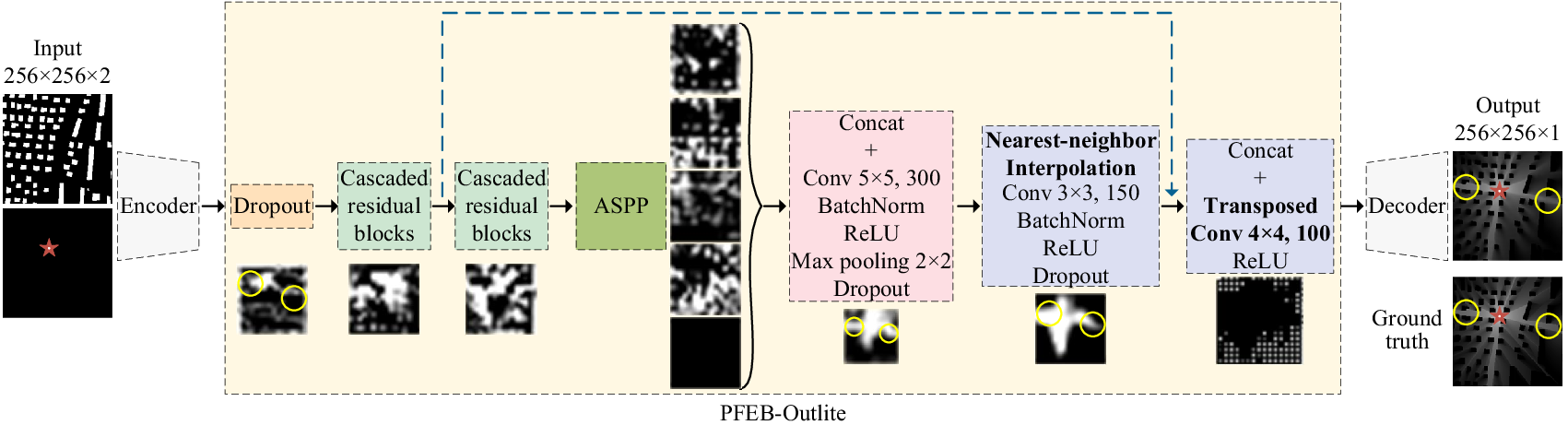}
  \caption{R$^{2}$Net-Outlite architecture.}
  \label{R2Net-Outlite}
  \vspace{-3mm}
\end{figure*}

\subsubsection{PFEB-Out}
In outdoor scenarios, the main pathloss comes from the diffraction loss brought by obstacles such as buildings~\cite{diffraction2,diffraction3}. The diffraction causes sudden signal attenuation and change of propagation direction, resulting in nonlinear impact on the outdoor radio map. To effectively extract nonlinear features, PFEB-Out adopts five cascaded residual blocks on $16\times16$ feature maps, which can estimate pathloss far away from the transmitter, as highlighted by the yellow circles in Fig.~\ref{R2Net-Out}. Moreover, buildings are of diverse sizes, which requires the network to be capable of extracting both local and global features. As such, ASPP with parallel atrous convolutions at multiple dilation rates is applied following the five cascaded residual blocks. As depicted in Fig.~\ref{R2Net-Out}, ASPP extracts features in different receptive fields, which are then fused by a $1\times1$ convolutional layer. It can be seen in Fig.~\ref{R2Net-Out} that compared to the input feature map of ASPP, the output feature map is more similar to the ground truth, especially in the circled areas where the receiver suffers from high diffraction loss. This showcases that the proposed PFEB-Out can well extract the features of diffraction loss.

To better fuse the features extracted by ASPP, PFEB-Out adopts transposed convolutions after ASPP to upsample the feature maps by zero padding. In addition, transposed convolution is adopted for all upsampling blocks in the decoder of R$^{2}$Net-Out with kernel size $6\times6$, as illustrated in Fig.~\ref{interpolation}(b). Moreover, after the first upsampling block in the decoder, as depicted in Fig.~\ref{R2Net}, R$^{2}$Net-Out employs a $5\times5$ convolutional layer. Here, the kernel size is larger than that of R$^{2}$Net-In for the sake of capturing global features in a larger receptive field.

\subsubsection{Numbers of Channels}
To extract abundant features, the numbers of channels in R$^{2}$Net-Out are set as $C_{\textup{1}}=64$, $C_{\textup{2}}=256$, $C_{\textup{3}}=512$, $C_{\textup{4}}=1024$, $C_{\textup{5}}=64$ and $C_{\textup{6}}=32$.

\vspace{-3mm}
\subsection{R$^{2}$Net-Outlite: A Light Variant of R$^{2}$Net-Out}
\label{sec-R2Net-Outlite}

It should be noted that to extract abundant features of diffraction loss, R$^{2}$Net-Out employs five cascaded residual blocks and has larger numbers of channels than those in R$^{2}$Net-In, which would lead to a large number of model parameters to train and thus high computational and storage costs. Therefore, a large training dataset is required to fully exploit the potentials of the proposed R$^{2}$Net-Out. However, in some situations, only a small training dataset is available, and the computing power and/or the storage space could be limited. To accommodate these cases, a light variant of R$^{2}$Net-Out, referred to as R$^{2}$Net-Outlite, is proposed by replacing PFEB-Out with PFEB-Outlite, which decreases the number of cascaded residual blocks and the numbers of channels. In addition, a max pooling layer and three dropout layers are added in R$^{2}$Net-Outlite to improve its generalization ability, as illustrated in Fig.~\ref{R2Net-Outlite}.

\subsubsection{PFEB-Outlite}
To maintain high accuracy in the case with a small training dataset, PFEB-Outlite adopts a dropout layer with a dropout probability of $0.3$ on $16\times16$ feature maps before upsampling to enhance the generalization ability of the model. Moreover, to reduce the computational and storage costs, PFEB-Outlite only adopts two cascaded residual blocks on $16\times16$ feature maps, which, however, results in less nonlinear features being captured. To compensate for this, a max pooling layer is adopted following the ASPP to increase the receptive field so as to better capture the diffraction loss incurred by the obstacles far away. As highlighted by the yellow circles in Fig.~\ref{R2Net-Outlite}, compared to the first feature map of PFEB-Outlite, the feature map after the dropout layer extracts the pathloss for the areas far away from the transmitter more accurately. Yet max pooling layer and the dropout layer could cause feature loss. To compensate for this, nearest-neighbor interpolation is adopted on $8\times8$ feature maps to keep rich details. As we can see in Fig.~\ref{R2Net-Outlite}, compared to the input feature map of nearest-neighbor interpolation, its output feature map is more similar to the ground truth.

\subsubsection{Numbers of Channels}
The numbers of channels, i.e., $C_{\textup{1}}$, $C_{\textup{2}}$, $C_{\textup{3}}$, $C_{\textup{4}}$, $C_{\textup{5}}$ and $C_{\textup{6}}$, in R$^{2}$Net-Outlite should be smaller than those in R$^{2}$Net-Out. For simplicity, they can be set the same as those in R$^{2}$Net-In.

\section{Experimental Results}
\label{Estimating Radio Maps}
Experimental results are presented in this section to evaluate the performance of the proposed R$^{2}$Net for both indoor and outdoor radio map estimation by comparing with the state-of-the-art benchmarks. The benchmarks chosen for comparison are detailed as follows:

\begin{itemize}
\item DPM~\cite{DPM}: DPM is the physical simulation method used to create our 3D indoor radio map dataset, and thus generates the ground truth.
\item RadioUNet~\cite{RadioUNet}: RadioUNet estimates 2D radio maps by cascading two U-Nets.
\item FadeNet~\cite{FadeNet}: FadeNet estimates 2D radio maps based on a variant of U-Net by adding a stride-1 convolution at both the input and the output sides.
\item RadioTrans~\cite{Transformerradionet}: RadioTrans estimates 2D radio maps by adopting Transformer based spread layers.
\item PPNet~\cite{PPNet}: PPNet employs SegNet to estimate 2D radio maps.
\end{itemize}
Note that although the benchmarks were originally designed for 2D radio map estimation, they can be extended to estimate 3D radio maps by adopting our proposed 2D height embedding method in Section~\ref{subsection Environmental Images} and applying the 3D-to-2D task transform presented in Section~\ref{Modeling of 3D Indoor Radio Map Estimation}.

\vspace{-3mm}
\subsection{Experimental Settings}
The experiments of 3D indoor radio map estimation are performed on our newly created dataset, 3DiRM3200, in which $3,200$ samples are randomly split into a training set of $2,560$ samples, a validation set of $320$ samples and a test set of $320$ samples. Note that the radio maps samples in these three sets correspond to different building layouts, furniture layouts and transmitter locations. The pathloss threshold is set as $(L_{\textup{thr}})_{\rm dB} = -127$\,dB.

For outdoor radio map estimation, due to the lack of publicly available 3D radio map dataset, we evaluate our method on the popularly used 2D outdoor radio map dataset RadioMapSeer~\cite{RadioUNet} as a special use case of our proposed 3D radio map estimation model. The RadioMapSeer dataset consists of $56,000$ simulated radio maps, which are randomly divided into a training set of $40,000$ samples, a validation set of $8,000$ samples and a test set of $8,000$ samples with non-overlapping environmental images. In RadioMapSeer, the buildings are at a height of $25$\,m, and the transmitters are at a height of $1.5$\,m. However, in the environmental images of RadioMapSeer, each pixel value only takes the value of $1$ or $0$ to indicate the existence of the building or the transmitter, while the impact of height is ignored. To reflect the effect of the heights of the buildings and the transmitters, the proposed 2D height embedding method is adopted.

Both our proposed R$^{2}$Net models and the benchmarks are trained on one Nvidia Geforce RTX 3090 with CUDA 11.4 and PyTorch 1.11.0. Adam~\cite{Adam} is adopted to optimize the parameters of the model with a learning rate of $10^{-4}$. Each model runs $50$ epochs. The batch size is $2$ for 3D indoor radio map estimation and $8$ for 2D outdoor radio map estimation. To alleviate overfitting, in the validation set, the model with the smallest MSE is picked out of $50$ epochs. Since our dataset, 3DiRM3200, is created on CPU according to DPM by using WinProp, for the convenience of comparing the inference speeds of DPM and other computer vision based methods, all methods are tested on Intel Core i5-11400F for 3D indoor radio map estimation. For 2D outdoor radio map estimation, all methods are tested on Nvidia Geforce RTX 3090. All presented radio maps have the same dimension of $256\times256$ pixels, while the pixel interval is $8$\,cm for indoor radio maps and $1$\,m for outdoor radio maps, respectively. The horizontal and vertical axes of the radio maps are omitted for presentation brevity.

\vspace{-3mm}
\subsection{Performance Metrics}
\subsubsection{Accuracy}
The estimation accuracy of radio maps is evaluated by the normalized mean square error (NMSE), the root mean square error (RMSE), the structural similarity (SSIM) and the peak signal-to-noise ratio (PSNR). NMSE is given by
\begin{equation}
    \text{NMSE}={\frac{1}{N_\textup{test}}}\sum_{i=1}^{N_\textup{test}}\frac{\lVert\widetilde{\mathbf{V}}(i)-{\mathbf{V}}(i)\rVert^2_2}{\lVert{\mathbf{V}}(i)\rVert^2_2},
    \label{eq:NMSE}
\end{equation}
where $N_\textup{test}$ is the size of test set. RMSE is the square root of MSE, defined as
\begin{equation}
    \text{RMSE} = \sqrt{{\frac{1}{N_\textup{test}}}\sum_{i=1}^{N_\textup{test}}\lVert\widetilde{\mathbf{V}}(i)-{\mathbf{V}}(i)\rVert^2_2}.
    \label{eq:RMSE}
\end{equation}
SSIM quantifies the structural similarity between the estimated radio map and the ground truth, which is expressed as
\begin{equation}
\text{SSIM}={\frac{1}{N_\textup{test}}}\sum_{i=1}^{N_\textup{test}}\frac{(2\mu_{{\mathbf{V}}(i)}\mu_{\widetilde{\mathbf{V}}(i)}+c_1)(2\sigma_{{\mathbf{V}}(i)\widetilde{\mathbf{V}}(i)} + c_2)}{(\mu_{{\mathbf{V}}(i)}^2+\mu_{\widetilde{\mathbf{V}}(i)}^2+c_1)(\sigma_{{\mathbf{V}}(i)}^2+\sigma_{\widetilde{\mathbf{V}}(i)}^2+c_2)},
\end{equation}
where $\mu_{{\mathbf{V}}(i)}$ and $\sigma_{{\mathbf{V}}(i)}^2$ are the mean and variance of the $i$-th ground truth, $\mu_{\widetilde{\mathbf{V}}(i)}$ and $\sigma_{\widetilde{\mathbf{V}}(i)}^2$ are the mean and variance of the $i$-th radio map estimate, and $\sigma_{{\mathbf{V}}(i)\widetilde{\mathbf{V}}(i)}$ is their covariance. $c_1=(k_1L)^2$ and $c_2=(k_2L)^2$ are the constants added to avoid the denominator being zero, where $L=1$ is the maximum difference between any two pixel values, and $k_1=0.01$ and $k_2=0.03$ are empirically chosen~\cite{SSIM}. SSIM takes a value between $-1$ and $1$, with $1$ indicating perfect structural similarity. PSNR measures pixel-level fidelity of radio maps, given by
\begin{equation}
\text{PSNR} = {\frac{1}{N_\textup{test}}}\sum_{i=1}^{N_\textup{test}}10 \cdot \log_{10} \left(\frac{\text{MAX}^2}{\lVert\widetilde{\mathbf{V}}(i)-{\mathbf{V}}(i)\rVert^2_2} \right),
\end{equation}
where $\text{MAX}$ is the maximum pixel value in all samples ($\text{MAX}=1$ in this paper). A higher PSNR indicates a higher image quality.

\subsubsection{Efficiency}
The efficiency of the proposed R$^{2}$Net is evaluated in terms of the following four metrics:
\begin{itemize}
\item Model size: Model size is the number of parameters (\#param). A large model size requires large memory storage and high computational cost for model training.
\item Multiply-accumulate operations (MACs): MACs are the number of multiplication and addition operations performed in the model, which affects both the computational cost and the inference speed.
\item Throughput: Throughput is defined as the number of radio maps estimated per second, i.e., the inference speed.
\item Inference time: Inference time is defined as the time required to obtain the radio map estimate per km$^{2}$.
\end{itemize}

\begin{table*}[t]
  \centering
  \caption{Comparison of R$^{2}$Net-In and the state-of-the-art benchmarks for 3D indoor radio map estimation. The best result is highlighted in bold, and * indicates the second best result.}
  \begin{tabular}{c|>{\centering\arraybackslash}p{1.5cm}>{\centering\arraybackslash}p{1.6cm}|>{\centering\arraybackslash}p{0.7cm}>{\centering\arraybackslash}p{0.7cm}>{\centering\arraybackslash}p{0.7cm}>{\centering\arraybackslash}p{0.7cm}
  |>{\centering\arraybackslash}p{0.7cm}>{\centering\arraybackslash}p{0.7cm}>{\centering\arraybackslash}p{1.1cm}>{\centering\arraybackslash}p{2cm}}
    \Xhline{1.0pt}
    \multirow{2}{*}{Methods} & \multirow{2}{*}{Training loss} & \multirow{2}{*}{Validation loss} & \multirow{2}{*}{NMSE} & \multirow{2}{*}{RMSE} & \multirow{2}{*}{SSIM} & \multirow{2}{*}{PSNR} & \multirow{2}{*}{\#param.} & \multirow{2}{*}{MACs} & \multirow{2}{*}{Throughput} & Inference time per km$^{2}$\\
    \hline
    DPM~\cite{DPM} & - & - & - & - & - & - & - & - & 0.04 & 3440.69\,s \\
    RadioUNet~\cite{RadioUNet} & \textbf{0.0003} & 0.0042* & 0.0292* & 0.0471* & 0.8883* & 29.54* & 13\,M* & 25.8\,G & 2* & 74.80\,s \\
    FadeNet~\cite{FadeNet} & 0.0020 & 0.0054 & 0.0583 & 0.0604 & 0.7723 & 26.25 & 65\,M & 51.7\,G & 1 & 84.15\,s \\
    RadioTrans~\cite{Transformerradionet} & 0.0021 & 0.0056 & 0.0714 & 0.0680 & 0.7228 & 25.44 & 55\,M & 18.7\,G* & 2* & 65.45\,s* \\
    PPNet~\cite{PPNet} & 0.0063 & 0.0091 & 0.1433 & 0.0935 & 0.5746 & 21.82 & 15\,M & 34.7\,G & 1 & 84.15\,s \\
    \hline
    R$^{2}$Net-In & 0.0010* & \textbf{0.0033} & \textbf{0.0268} & \textbf{0.0395} & \textbf{0.8908} & \textbf{29.65} & \textbf{8\,M} & \textbf{6.5\,G} & \textbf{3} & \textbf{46.75\,s} \\
    \Xhline{1.0pt}
  \end{tabular}
  \label{Comparision of the RadioResNet-3D and previous state-of-the-art}
    \vspace{-3mm}
\end{table*}

\vspace{-3mm}
\subsection{Experimental Results of 3D Indoor Radio Map Estimation}
\subsubsection{Quantitative Results}
Table~\ref{Comparision of the RadioResNet-3D and previous state-of-the-art} lists the training loss, validation loss, NMSE, RMSE, SSIM, PSNR, model size (\#param), MACs, throughput and inference time per km$^{2}$ of the proposed R$^{2}$Net-In and the state-of-the-art benchmarks for 3D indoor radio map estimation. It can be seen that R$^{2}$Net-In achieves the smallest validation loss and the second smallest training loss, which indicates that R$^{2}$Net-In has strong generalization ability. The training loss of RadioUNet~\cite{RadioUNet} is the smallest, however, it has larger validation loss than R$^{2}$Net-In, implying that the generalization ability of RadioUNet is not as good as that of R$^{2}$Net-In. The training loss and the validation loss of the other benchmarks are neither the best nor the second best, indicating that the other benchmarks are underfitting. In addition, we can see that R$^{2}$Net-In achieves the smallest NMSE, which is $8.22\%$ smaller than that of RadioUNet, and $81.30\%$ smaller than that of PPNet~\cite{PPNet}. The RMSE of R$^{2}$Net-In is also the smallest, which is $16.14\%$ smaller than that of RadioUNet. R$^{2}$Net-In also achieves the highest SSIM and PSNR, indicating that the radio map estimates obtained by our method share the highest structural similarity with the ground truth and have the highest quality. The above quantitative results validate that our R$^{2}$Net-In can notably improve the estimation accuracy of 3D indoor radio maps.

As regards to the efficiency, R$^{2}$Net-In is roughly two orders of magnitude faster than DPM~\cite{DPM}, highlighting the importance of developing deep learning based computer vision methods for radio map estimation. Moreover, our model involves much less MAC operations and has higher throughput and shorter inference time than the benchmarks, verifying its computational efficiency. In addition, the model size of the proposed R$^{2}$Net-In is significantly smaller than the benchmarks. Specifically, it is $38.46\%$ smaller than RadioUNet~\cite{RadioUNet} and $87.69\%$ smaller than FadeNet~\cite{FadeNet}. This indicates that R$^{2}$Net-In occupies less amount of memory storage than the benchmarks.

\begin{table}[t]
  \centering
  \caption{Detailed NMSE on test samples. The best result is highlighted in bold, and * indicates the second best result.}
  \begin{tabular}{c|cccc}
    \Xhline{1.0pt}
    \multirow{2}*{Methods} &
    \multicolumn{4}{c}{NMSE}\\
    \cline{2-5}
     & Min & Max & Average & 95\% CI \\
    \hline
    RadioUNet~\cite{RadioUNet} & \textbf{0.0043} & \textbf{0.5250} & 0.0292* & [\textbf{0.0060}, 0.1469*] \\
    FadeNet~\cite{FadeNet} & 0.0095 & 0.6103* & 0.0583 & [0.0150, 0.2524] \\
    RadioTrans~\cite{Transformerradionet} & 0.0181 & 0.9247 & 0.0714 & [0.0211, 0.2604] \\
    PPNet~\cite{PPNet} & 0.0311 & 0.7457 & 0.1433 & [0.0385, 0.4080] \\
    \hline
    R$^{2}$Net-In & 0.0050* & 0.6274 & \textbf{0.0268} & [0.0064*, \textbf{0.0768}] \\
    \Xhline{1.0pt}
  \end{tabular}
  \label{Accuracy of the RadioResNet-3D and previous state-of-the-art}
    \vspace{-3mm}
\end{table}

\begin{figure*}[t]
\centering
\captionsetup[subfloat]{labelsep=none,format=plain,labelformat=empty}
\subfloat[]{\begin{minipage}{0.03\linewidth}\vspace{-65pt}\hspace{-5pt}\scriptsize{$0.5$\,m}\end{minipage}}
\subfloat[]{\includegraphics[width=0.0674\linewidth]{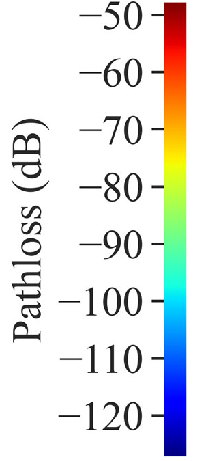}}
\subfloat[]{\includegraphics[width=0.15\linewidth]{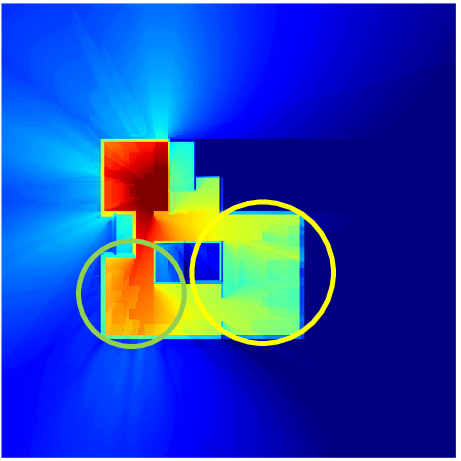}}
\subfloat[]{\includegraphics[width=0.15\linewidth]{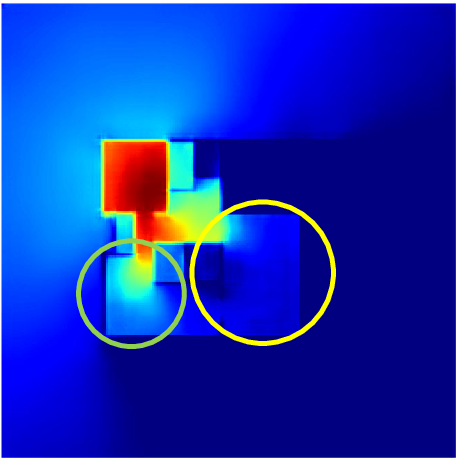}}
\subfloat[]{\includegraphics[width=0.15\linewidth]{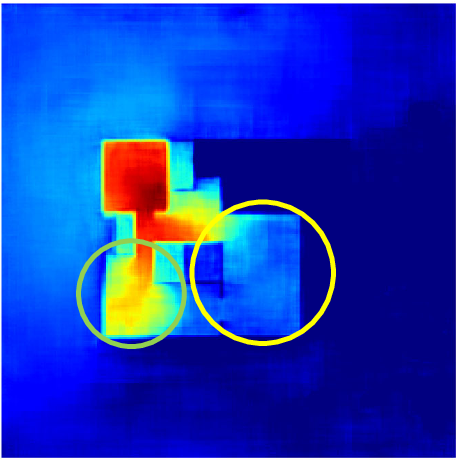}\label{fadenet}}
\subfloat[]{\includegraphics[width=0.15\linewidth]{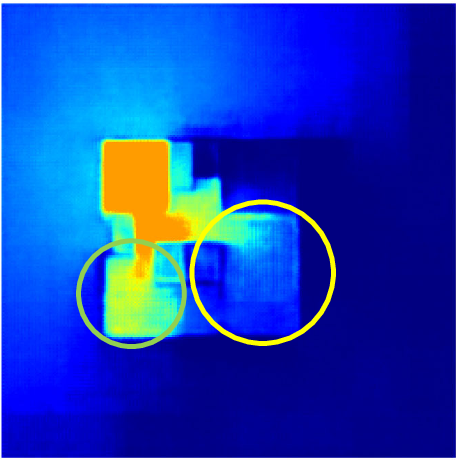}}
\subfloat[]{\includegraphics[width=0.15\linewidth]{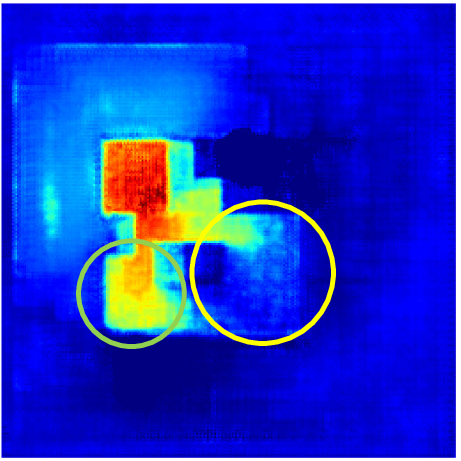}}
\subfloat[]{\includegraphics[width=0.15\linewidth]{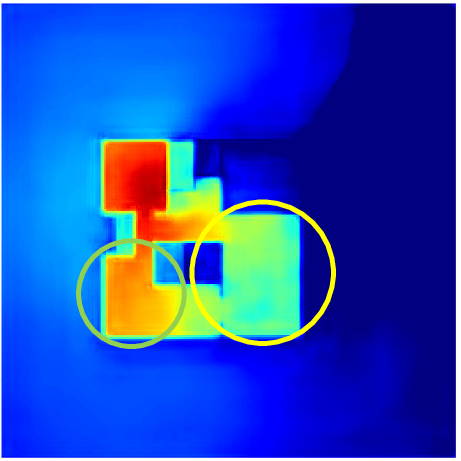}}\\\vspace{-2.1em}
\subfloat[]{\begin{minipage}{0.03\linewidth}\vspace{-65pt}\hspace{-5pt}\scriptsize{$1$\,m}\end{minipage}}
\subfloat[]{\includegraphics[width=0.0674\linewidth]{fig/colorbar_197_3.eps}}
\subfloat[]{\includegraphics[width=0.15\linewidth]{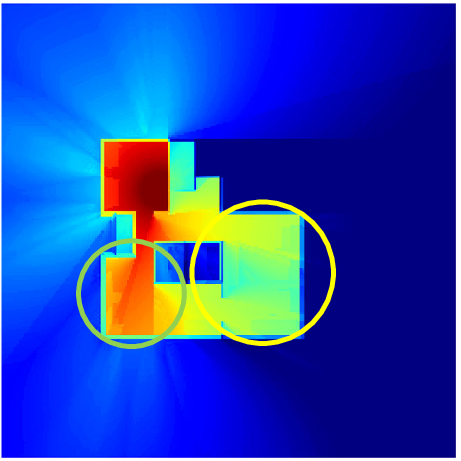}}
\subfloat[]{\includegraphics[width=0.15\linewidth]{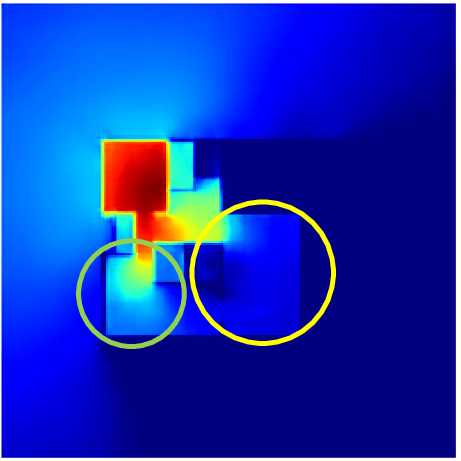}}
\subfloat[]{\includegraphics[width=0.15\linewidth]{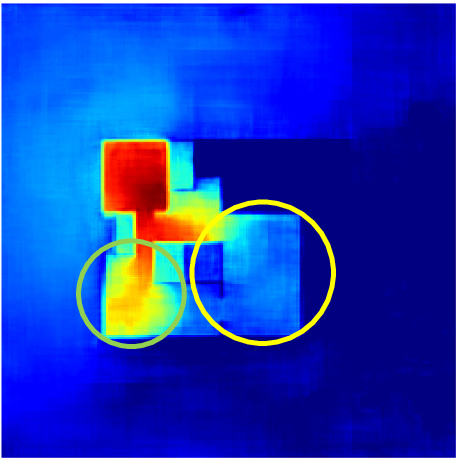}\label{fadenet}}
\subfloat[]{\includegraphics[width=0.15\linewidth]{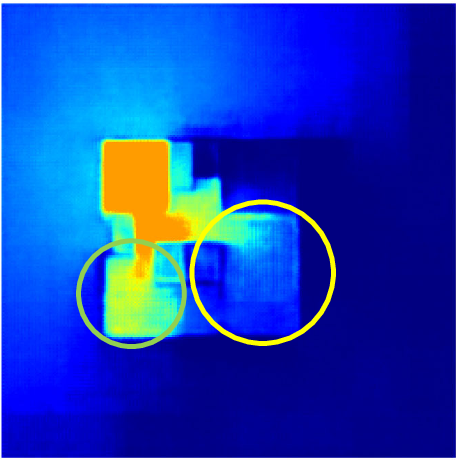}}
\subfloat[]{\includegraphics[width=0.15\linewidth]{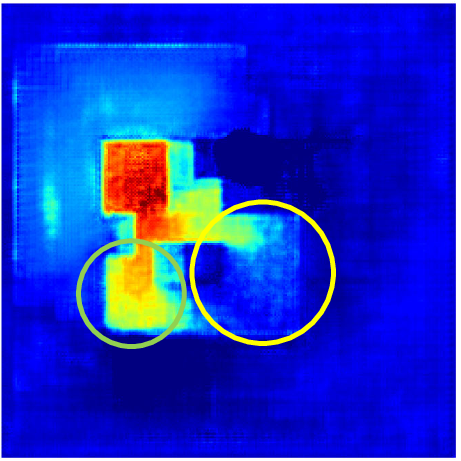}}
\subfloat[]{\includegraphics[width=0.15\linewidth]{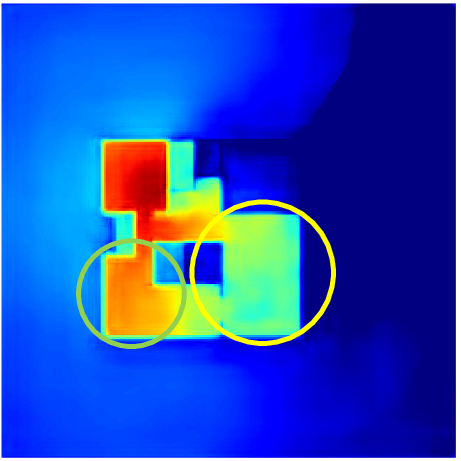}}\\\vspace{-2.1em}
\subfloat[]{\begin{minipage}{0.03\linewidth}\vspace{-65pt}\hspace{-5pt}\scriptsize{$1.5$\,m}\end{minipage}}
\subfloat[]{\includegraphics[width=0.0674\linewidth]{fig/colorbar_197_3.eps}}
\subfloat[(a) Ground truth]{\includegraphics[width=0.15\linewidth]{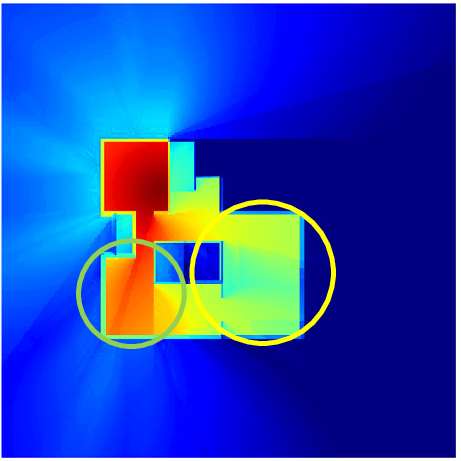}}
\subfloat[(b) RadioUNet~\cite{RadioUNet}]{\includegraphics[width=0.15\linewidth]{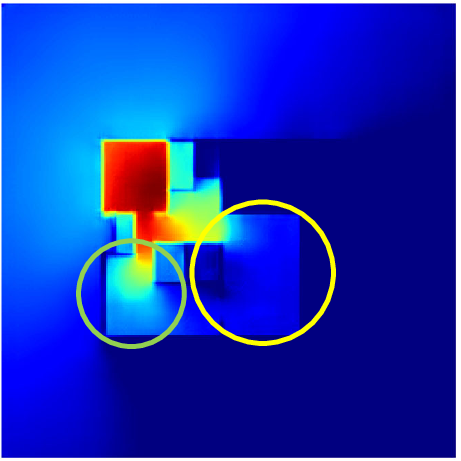}}
\subfloat[(c) FadeNet~\cite{FadeNet}]{\includegraphics[width=0.15\linewidth]{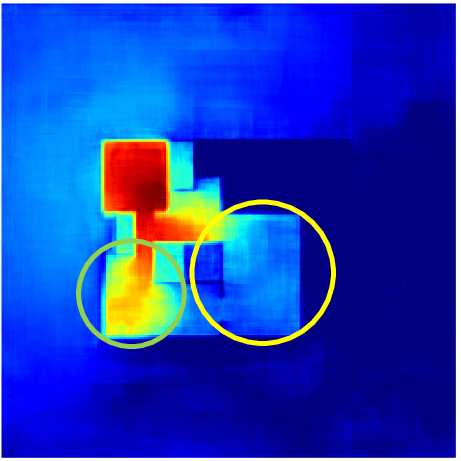}\label{fadenet}}
\subfloat[(d) RadioTrans~\cite{Transformerradionet}]{\includegraphics[width=0.15\linewidth]{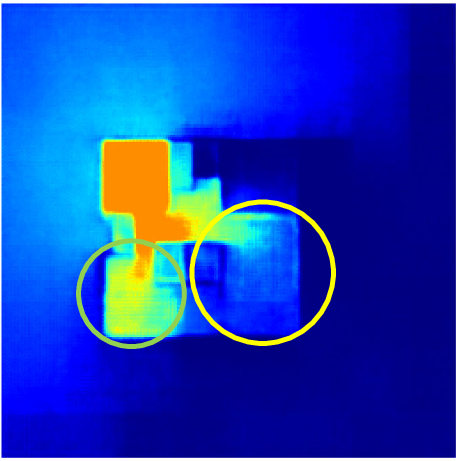}}
\subfloat[(e) PPNet~\cite{PPNet}]{\includegraphics[width=0.15\linewidth]{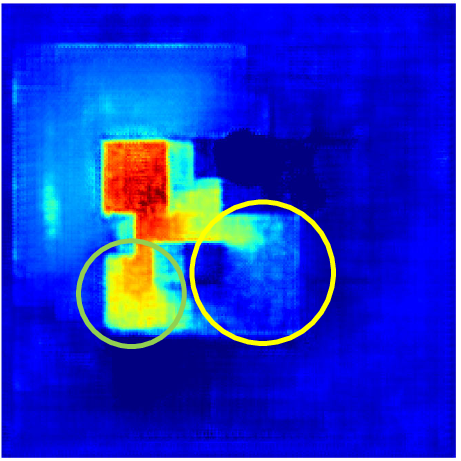}}
\subfloat[(f) R$^{2}$Net-In]{\includegraphics[width=0.15\linewidth]{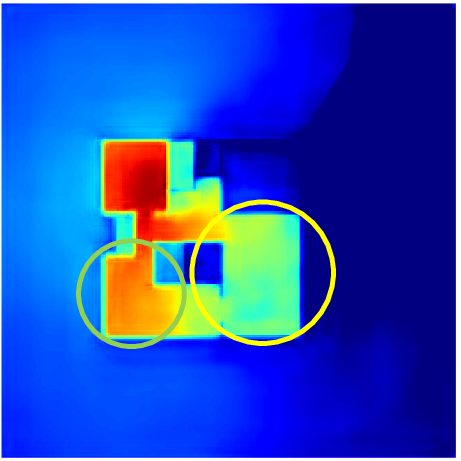}}
\caption{Visualization of estimation results of R$^{2}$Net-In and the state-of-the-art benchmarks at heights of $0.5$\,m, $1$\,m and $1.5$\,m for a randomly chosen 3D indoor test sample.}
\label{Comparision of the RadioResNet3d}
  \vspace{-3mm}
\end{figure*}

To evaluate the overall accuracy of the proposed R$^{2}$Net-In on all test samples, Table~\ref{Accuracy of the RadioResNet-3D and previous state-of-the-art} takes a closer look at the achieved NMSE by showing the minimum NMSE, the maximum NMSE, the average NMSE and the $95\%$ confidence interval. It can be seen that the confidence interval of our R$^{2}$Net-In is the narrowest. In addition, R$^{2}$Net-In has the smallest upper-bound of the $95\%$ confidence interval among all models. Both of the above observations corroborate that our R$^{2}$Net-In is more robust than the state-of-the-art benchmarks and can accurately estimate the radio maps in various cases. Particularly, the NMSE scores with our R$^{2}$Net-In are mainly distributed between $0.0064$ and $0.0768$, while those with RadioUNet are distributed between $0.0060$ and $0.1469$. As the upper-bound of the $95\%$ confidence interval with R$^{2}$Net-In is $47.72\%$ less than that with RadioUNet, our R$^{2}$Net-In estimates radio maps more accurately than RadioUNet for most samples. 

\subsubsection{Qualitative Results}
Fig.~\ref{Comparision of the RadioResNet3d} further visualizes the predicted radio maps at heights of $0.5$\,m, $1$\,m and $1.5$\,m for a randomly chosen test sample. We can clearly see from the figures that the radio maps obtained from R$^{2}$Net-In are very close to the ground truth. In particular, R$^{2}$Net-In can still well predict the pathloss far away from the transmitter as highlighted by the yellow circles, while the benchmarks fail, corroborating that the proposed R$^{2}$Net-In can successfully capture complex indoor pathloss features. This is because we adopt dropout layers to improve the generalization ability of the model and adopt cascaded residual blocks to better capture nonlinear features. In addition, with an increase in height, the area highlighted by the green circle in the ground truth appears darker due to the existence of furniture at comparable heights. This phenomenon can only be successfully captured by our R$^{2}$Net-In, as can be observed in Fig.~\ref{Comparision of the RadioResNet3d}, which further validates the effectiveness of the proposed R$^{2}$Net-In in 3D indoor radio map estimation.

\begin{table}[t]
  \centering
  \caption{Ablation on dropout layers, residual blocks and height embedding for R$^{2}$Net-In. The best result is highlighted in bold.}
  \begin{tabular}{c|cccc}
    \Xhline{1.0pt}
    Methods & NMSE & RMSE & SSIM & PSNR \\
    \hline
    w/o dropout layers & 0.0361 & 0.0467 & 0.8634 & 29.53 \\
    w/o residual blocks & 0.0369 & 0.0487 & 0.8661 & 29.62 \\
    w/o height embedding & 0.0328 & 0.0451 & 0.8841 & 29.60 \\
    \hline
    R$^{2}$Net-In & \textbf{0.0268} & \textbf{0.0395} & \textbf{0.8908} & \textbf{29.65} \\
    \Xhline{1.0pt}
  \end{tabular}
  \label{Ablation indoor}
\end{table}

\begin{table*}[t]
  \centering
  \caption{Comparison of R$^{2}$Net-Out, R$^{2}$Net-Outlite and the state-of-the-art benchmarks for 2D outdoor radio map estimation. The best result is highlighted in bold, and * indicates the second best result.}
  \begin{tabular}{c|>{\centering\arraybackslash}p{1.5cm}>{\centering\arraybackslash}p{1.6cm}|>{\centering\arraybackslash}p{0.7cm}>{\centering\arraybackslash}p{0.7cm}>{\centering\arraybackslash}p{0.7cm}>{\centering\arraybackslash}p{0.7cm}
  |>{\centering\arraybackslash}p{0.7cm}>{\centering\arraybackslash}p{0.7cm}>{\centering\arraybackslash}p{1.1cm}>{\centering\arraybackslash}p{2cm}}
    \Xhline{1.0pt}
    \multirow{2}{*}{Methods} & \multirow{2}{*}{Training loss} & \multirow{2}{*}{Validation loss} & \multirow{2}{*}{NMSE} & \multirow{2}{*}{RMSE} & \multirow{2}{*}{SSIM} & \multirow{2}{*}{PSNR} & \multirow{2}{*}{\#param.} & \multirow{2}{*}{MACs} & \multirow{2}{*}{Throughput} & Inference time per km$^{2}$\\
    \hline
    RadioUNet~\cite{RadioUNet} & \textbf{0.0001} & 0.0004* & 0.0106 & 0.0202 & 0.9202 & 34.43 & 13\,M* & 23.5\,G & 127 & 0.1190\,s \\
    FadeNet~\cite{FadeNet} & 0.0006 & 0.0006 &0.0142 & 0.0240 & 0.9023 & 32.83 & 65\,M & 49.1\,G & 105 & 0.1450\,s \\
    RadioTrans~\cite{Transformerradionet} & \textbf{0.0001} & 0.0004* & 0.0101 & 0.0196 & 0.9267 & 34.57 & 55\,M & 18.7\,G* & 144* & 0.1053\,s* \\
    PPNet~\cite{PPNet} & 0.0031 & 0.0040 & 0.0986 & 0.0647 & 0.7417 & 24.39 & 15\,M & 34.0\,G & \textbf{198} & \textbf{0.0778\,s} \\
    \hline
    R$^{2}$Net-Out & \textbf{0.0001} & \textbf{0.0003} & \textbf{0.0046} & \textbf{0.0156} & \textbf{0.9491} & \textbf{37.00} & 85\,M & 161.0\,G & 71 & 0.2136\,s \\
    R$^{2}$Net-Outlite & 0.0002* & \textbf{0.0003} & 0.0059* & 0.0178* & 0.9403* & 35.92* & \textbf{4\,M} & \textbf{13.8\,G} & 115 & 0.1328\,s \\
    \Xhline{1.0pt}
  \end{tabular}
  \label{R2Net and state-of-the-arts on 2D outdoor radio map estimation}
    \vspace{-3mm}
\end{table*}

\subsubsection{Ablation Study}
To demonstrate the effectiveness of dropout layers, residual blocks and height embedding adopted in our R$^{2}$Net-In, Table~\ref{Ablation indoor} lists the NMSE, RMSE, SSIM and PSNR of the proposed R$^{2}$Net-In and its ablation versions for 3D indoor radio map estimation by removing all dropout layers, replacing the residual blocks with canonical convolutional layers, or setting the pixel value as $1$ or $0$ to indicate the existence of an object without height embedding. We can see from the table that removing dropout layers, residual blocks or height embedding increases NMSE and RMSE, and decreases SSIM and PSNR. This validates the effectiveness of all these three components in improving the accuracy of indoor radio map estimation, among which dropout layers and residual blocks contribute more compared to height embedding. This is because in indoor scenarios, the height difference across various objects is usually limited to a small range, leading to less effects on radio map estimation.

\vspace{-3mm}
\subsection{Experimental Results\hspace{-0.2mm} of \hspace{-0.2mm}2D Outdoor Radio Map Estimation}
\subsubsection{Quantitative Results}
Table~\ref{R2Net and state-of-the-arts on 2D outdoor radio map estimation} lists the training loss, validation loss, NMSE, RMSE, SSIM, PSNR, model size (\#param), MACs, throughput and inference time per km$^{2}$ of the proposed R$^{2}$Net-Out and R$^{2}$Net-Outlite, and the state-of-the-art benchmarks for 2D outdoor radio map estimation. As can be seen from the table, R$^{2}$Net-Out and R$^{2}$Net-Outlite achieve the smallest validation loss than the benchmarks, which indicates that they have better generalization ability. Both R$^{2}$Net-Out and R$^{2}$Net-Outlite achieve smaller NMSE and RMSE, and higher SSIM and PSNR than the benchmarks. Particularly, R$^{2}$Net-Out achieves the smallest NMSE, which is $44.34\%$ less than that of RadioUNet~\cite{RadioUNet}, and $95.33\%$ less than that of PPNet~\cite{PPNet}. The RMSE of R$^{2}$Net-Out is also the smallest, which is $20.41\%$ smaller than that of RadioTrans~\cite{Transformerradionet}. In addition, R$^{2}$Net-Out also achieves the highest SSIM and PSNR, which validates the superior performance of the proposed method in preserving both structural information and pixel-level fidelity of the radio maps over the state-of the-art. The above results show that the proposed R$^{2}$Net-Out can significantly improve the accuracy of 2D outdoor radio map estimation, yet at the cost of larger model size, more MACs, and longer inference time. By contrast, the model size of R$^{2}$Net-Outlite is the smallest, which is $69.23\%$ smaller than that of RadioUNet, and $93.85\%$ smaller than FadeNet~\cite{FadeNet}. R$^{2}$Net-Outlite also involves less MACs than the benchmarks, and it estimates 2D outdoor radio maps faster than FadeNet. These results clearly indicate that R$^{2}$Net-Outlite can estimate 2D outdoor radio maps efficiently than the benchmarks, while maintaining good estimation accuracy.

\begin{figure*}[t]
\centering
\captionsetup[subfloat]{labelsep=none,format=plain,labelformat=empty}
\subfloat[]{\includegraphics[width=0.067\linewidth]{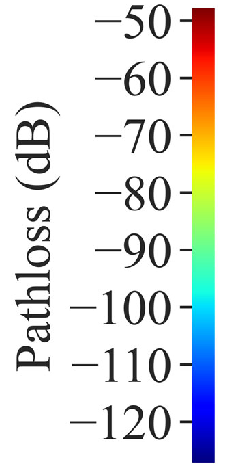}}
\subfloat[(a) Ground truth]{\includegraphics[width=0.13\linewidth]{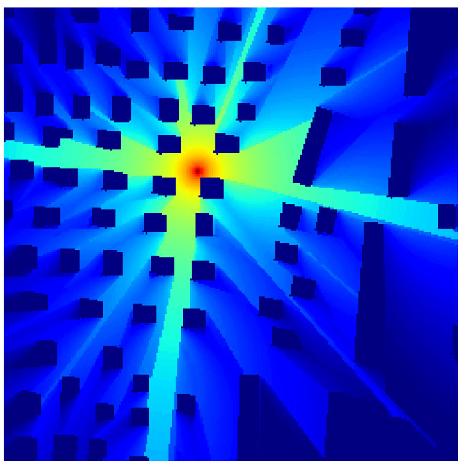}}
\subfloat[(b) RadioUNet~\cite{RadioUNet}]{\includegraphics[width=0.13\linewidth]{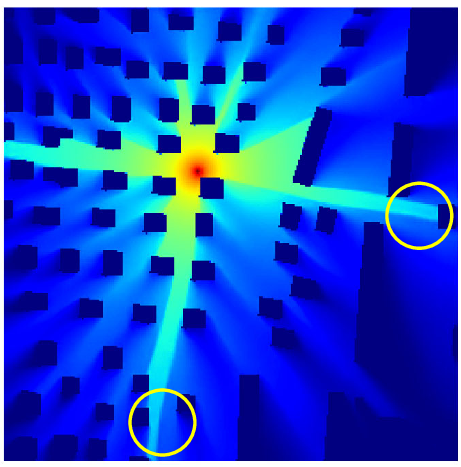}}
\subfloat[(c) FadeNet~\cite{FadeNet}]{\includegraphics[width=0.13\linewidth]{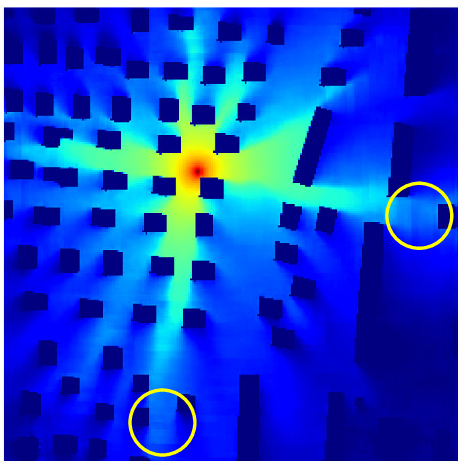}}
\subfloat[(d) RadioTrans~\cite{Transformerradionet}]{\includegraphics[width=0.1305\linewidth]{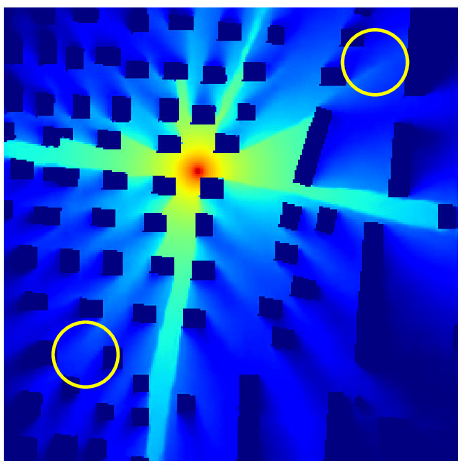}}
\subfloat[(e) PPNet~\cite{PPNet}]{\includegraphics[width=0.13\linewidth]{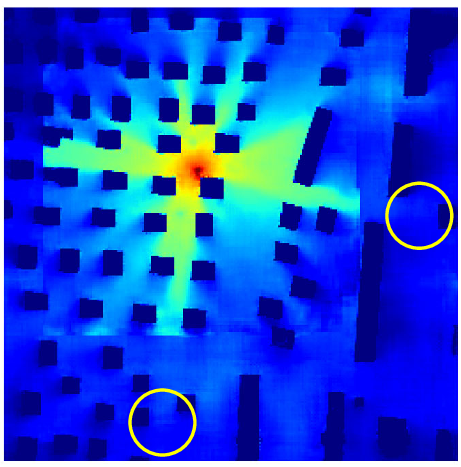}}
\subfloat[(f) R$^{2}$Net-Out]{\includegraphics[width=0.13\linewidth]{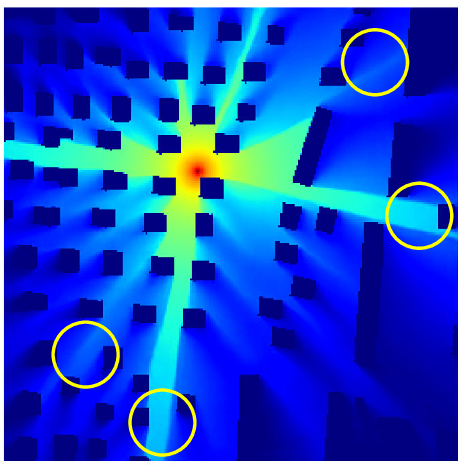}}
\subfloat[(g) R$^{2}$Net-Outlite]{\includegraphics[width=0.13\linewidth]{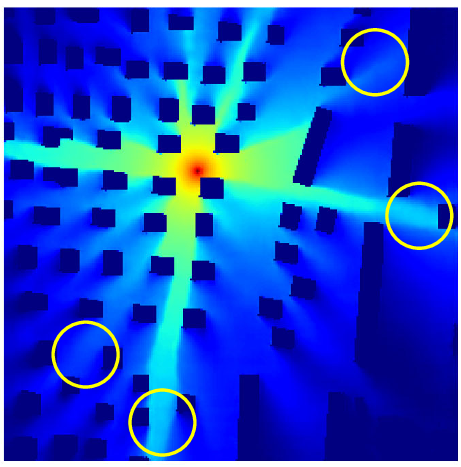}}
\caption{Visualization of estimation results of R$^{2}$Net-Out, R$^{2}$Net-Outlite and the state-of-the-art benchmarks for a randomly chosen 2D outdoor test sample.}
\label{Visualization of the R2Net on 2D outdoor radio map estimation}
  \vspace{-3mm}
\end{figure*}

\subsubsection{Qualitative Results}
Fig.~\ref{Visualization of the R2Net on 2D outdoor radio map estimation} visualizes the estimated radio maps for a randomly chosen test sample. It is clear that the radio maps estimated by both R$^{2}$Net-Out and R$^{2}$Net-Outlite are very close to the ground truth. Particularly, when the receiver is far from the transmitter, R$^{2}$Net-Out and R$^{2}$Net-Outlite can still predict the pathloss accurately, while the benchmarks are incapable. This evidently verifies the ability of R$^{2}$Net-Out and R$^{2}$Net-Outlite to enhance the feature extraction of outdoor pathloss.

\begin{table}[t]
  \centering
  \caption{Accuracy of R$^{2}$Net-Out, R$^{2}$Net-Outlite and the state-of-the-art benchmarks on small datasets for 2D outdoor radio map estimation. The best result is highlighted in bold, and * indicates the second best result.}
  \begin{tabular}{c|ccccc}
    \Xhline{1.0pt}
    \multirow{2}*{Methods} &
    \multicolumn{5}{c}{NMSE}\\
    \cline{2-6}
    & Tx16 & Tx32 & Tx48 & Tx64 & Tx80 \\
    \hline
    RadioUNet~\cite{RadioUNet} & 0.0269 & 0.0145* & 0.0130* & 0.0119 & 0.0120 \\
    FadeNet~\cite{FadeNet} & 0.0393 & 0.0254 & 0.0171 & 0.0186 & 0.0157 \\
    RadioTrans~\cite{Transformerradionet} & 0.0317 & 0.0187 & 0.0174 & 0.0151 & 0.0124 \\
    PPNet~\cite{PPNet} & 0.1071 & 0.1012 & 0.0992 & 0.0983 & 0.0975 \\
    \hline
    R$^{2}$Net-Out & 0.0236* & \textbf{0.0135} & \textbf{0.0114} & \textbf{0.0099} & \textbf{0.0085} \\
    R$^{2}$Net-Outlite & \textbf{0.0228}& 0.0152 & 0.0140 & 0.0117* & 0.0109* \\
    \Xhline{1.0pt}
  \end{tabular}
  \label{Accuracy of the R2Net small}
    \vspace{-2mm}
\end{table}

\begin{table}[t]
  \centering
  \caption{Effectiveness of the proposed 2D height embedding method. The best result is highlighted in bold.}
  \begin{tabular}{c|cccc}
    \Xhline{1.0pt}
    NMSE & \makecell[c]{RadioUNet\\\cite{RadioUNet}} & \makecell[c]{FadeNet\\\cite{FadeNet}} & \makecell[c]{RadioTrans\\\cite{Transformerradionet}} & \makecell[c]{PPNet\\\cite{PPNet}} \\
    \hline
    w/o height embedding & 0.0106 & 0.0142 & 0.0101 & 0.0986 \\
    w/ height embedding & \textbf{0.0063} & \textbf{0.0096} & \textbf{0.0066} & \textbf{0.0779} \\
    \Xhline{1.0pt}
  \end{tabular}
  \label{Comparison of 2D outdoor radio map estimation with/without the proposed representation method}
    \vspace{-3mm}
\end{table}

\begin{figure}[t]
\centering
\captionsetup[subfloat]{labelsep=none,format=plain,labelformat=empty}
\subfloat[]{\includegraphics[width=0.1236\linewidth]{fig/colorbar_684_55_24.eps}}
\subfloat[(a) Ground truth]{\includegraphics[width=0.24\linewidth]{fig/target_684_55.eps}}
\\
\subfloat[(b) RadioUNet~\cite{RadioUNet}]{\includegraphics[width=0.2408\linewidth]{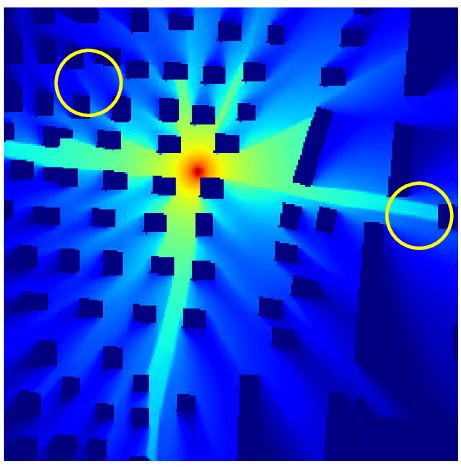}}
\subfloat[(c) RadioUNet~\cite{RadioUNet} w/ height embedding]{\includegraphics[width=0.24\linewidth]{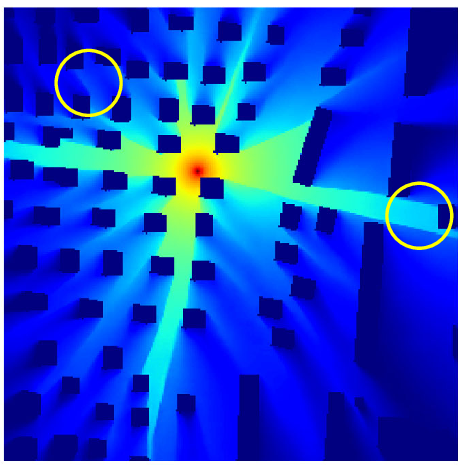}}
\subfloat[(d) FadeNet~\cite{FadeNet}]{\includegraphics[width=0.241\linewidth]{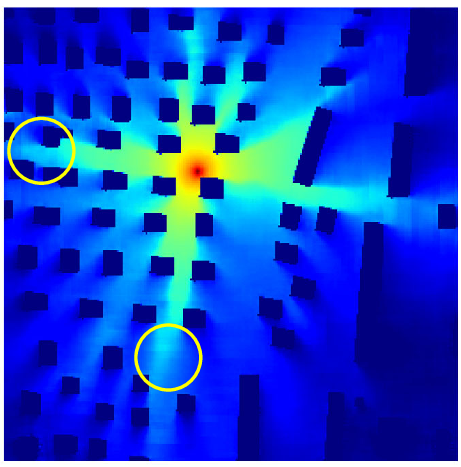}}
\subfloat[(e) FadeNet~\cite{FadeNet} w/ height embedding]{\includegraphics[width=0.2402\linewidth]{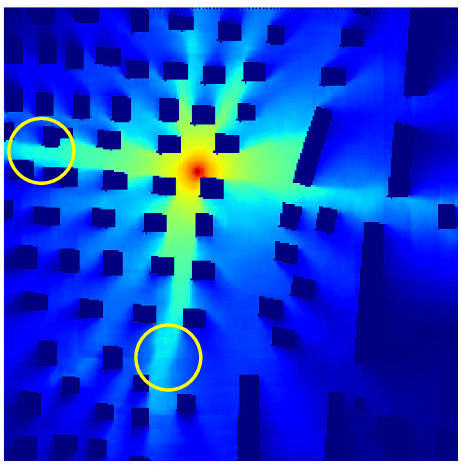}}
\\
\subfloat[(f) RadioTrans~\cite{Transformerradionet}]{\includegraphics[width=0.24\linewidth]{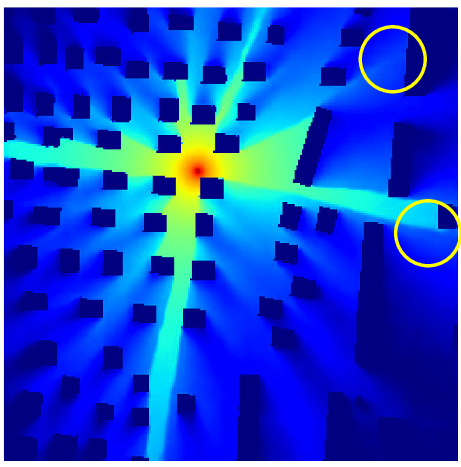}}
\subfloat[(g) RadioTrans~\cite{Transformerradionet} w/ height embedding]{\includegraphics[width=0.24\linewidth]{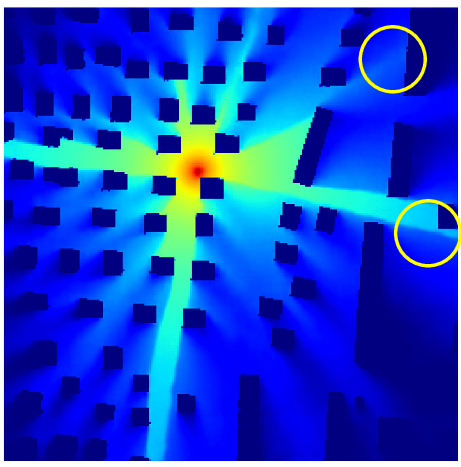}}
\subfloat[(h) PPNet~\cite{PPNet}]{\includegraphics[width=0.241\linewidth]{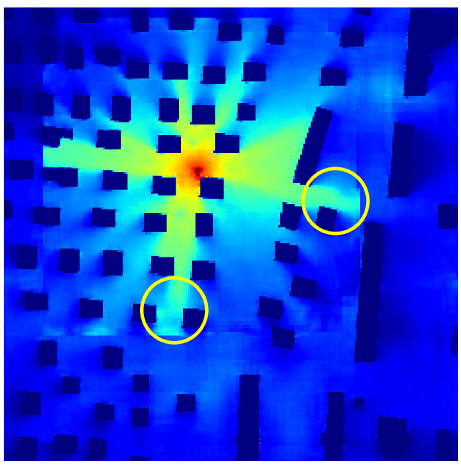}}
\subfloat[(i) PPNet~\cite{PPNet} w/ height embedding]{\includegraphics[width=0.2402\linewidth]{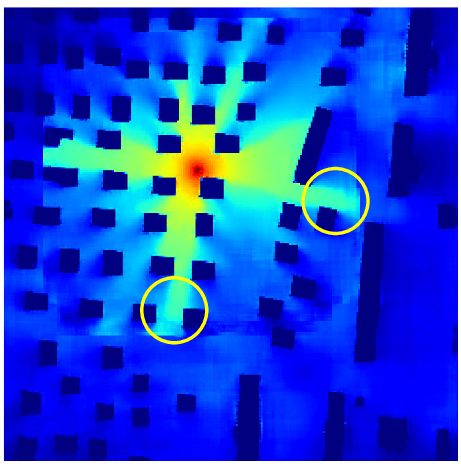}}
\caption{Visualization of the effectiveness of the proposed 2D height embedding method.}
\label{Visualization of the state-of-the-art benchmarks with/without the proposed representation method}
  \vspace{-3mm}
\end{figure}

\subsubsection{Comparison of R$^{2}$Net-Out and R$^{2}$Net-Outlite}
To investigate the impact of dataset size, $400$ city images are chosen randomly from RadioMapSeer with the corresponding $32,000$ radio maps and $80$ transmitter locations. The $400$ city images are randomly split into a training set of $320$ city images, a validation set of $40$ city images and a test set of $40$ city images. Furthermore, the number of transmitters per map is chosen as $16$, $32$, $48$, $64$ and $80$, respectively, in the training sets Tx16, Tx32, Tx48, Tx64 and Tx80. The smallest training set is $320$ city images with $16$ transmitters per map, leading to $5,120$ radio maps. As presented in Table~\ref{Accuracy of the R2Net small}, when the training set is the smallest (Tx16), R$^{2}$Net-Outlite achieves the smallest NMSE, which indicates that R$^{2}$Net-Outlite entails strong generalization ability and is more suitable for a small training dataset. When the number of transmitters is larger than $16$, R$^{2}$Net-Out achieves the smallest NMSE, which shows that R$^{2}$Net-Out should be chosen for accurate estimation when a large training dataset is available.

\subsubsection{Effectiveness of 2D Height Embedding}
To demonstrate the advantage of the proposed 2D height embedding method, Table~\ref{Comparison of 2D outdoor radio map estimation with/without the proposed representation method} lists the NMSE of the state-of-the-art benchmarks both with and without the 2D height embedding method. It can be seen that all the state-of-the-art benchmarks with the 2D height embedding method outperform those without 2D height embedding. Notably, the accuracy of RadioUNet is improved by $16\%$, and RadioTrans achieves $34.65\%$ higher accuracy. Furthermore, the visualization results are shown in Fig.~\ref{Visualization of the state-of-the-art benchmarks with/without the proposed representation method}. It is clear that when the receiver is far from the transmitter, the models that employ our 2D height embedding method can predict the pathloss more accurately.

\subsubsection{Experimental Results on Real-World Measured Dataset}
To further demonstrate the performance of R$^{2}$Net-Out in real-world urban environments, the experiments of 2D outdoor radio map estimation are also conducted on the real-world measured dataset, RSRPSet\_urban~\cite{RSRPSeturban}, which contains the reference signal receiving power (RSRP) measured by Huawei Technologies Co. Ltd in $180$ dense urban communication cells, equivalent to $180$ 2D outdoor radio maps. The $180$ radio maps are randomly divided into a training set of $150$ samples and a test set of $30$ samples. Table~\ref{R2Net and state-of-the-arts on the real-world measured dataset} lists the NMSE, RMSE, SSIM and PSNR of the proposed R$^{2}$Net-Out and R$^{2}$Net-Outlite, and the state-of-the-art benchmarks for 2D outdoor radio map estimation. As can be seen from the table, R$^{2}$Net-Out and R$^{2}$Net-Outlite achieve smaller NMSE and RMSE, and higher SSIM and PSNR than the benchmarks, which showcases the superior performance of our R$^{2}$Net-Out and R$^{2}$Net-Outlite in estimating outdoor radio maps in practice.

\begin{table}[t]
  \centering
  \caption{Comparison of R$^{2}$Net-Out, R$^{2}$Net-Outlite and the state-of-the-art benchmarks for 2D outdoor radio map estimation on the real-world measured dataset RSRPSet\_urban. The best result is highlighted in bold, and * indicates the second best result.}
  \begin{tabular}{c|cccc}
    \Xhline{1.0pt}
    Methods & NMSE & RMSE & SSIM & PSNR \\
    \hline
    RadioUNet~\cite{RadioUNet} & 0.2985 & 0.2650 & 0.4473 & 11.94 \\
    RadioTrans~\cite{Transformerradionet} & 0.3001 & 0.2645 & 0.4447 & 11.92 \\
    PPNet~\cite{PPNet} & 0.3001 & 0.2646 & 0.4469 & 11.94 \\
    \hline
    R$^{2}$Net-Out & \textbf{0.1834} & \textbf{0.2063} & \textbf{0.5275} & \textbf{13.98} \\
    R$^{2}$Net-Outlite & 0.1861* & 0.2092* & 0.5149* & 13.87* \\
    \Xhline{1.0pt}
  \end{tabular}
  \label{R2Net and state-of-the-arts on the real-world measured dataset}
\end{table}

\subsubsection{Ablation Study}
To demonstrate the effectiveness of ASPP, residual blocks and height embedding adopted in the proposed R$^{2}$Net-Out, Table~\ref{Ablation outdoor} presents the NMSE, RMSE, SSIM and PSNR of the proposed R$^{2}$Net-Out and its ablation versions. As can be seen from the table, in contrast to indoor radio map estimation, height embedding contributes the most to the accuracy improvement of outdoor radio map estimation. In fact, in outdoor scenarios, pathloss closely depends on diffraction loss. As diffraction loss could be significantly affected by building heights, height embedding plays a vital role in outdoor radio map estimation.

\begin{table}[t]
  \centering
  \caption{Ablation on ASPP, residual blocks and height embedding for R$^{2}$Net-Out. The best result is highlighted in bold.}
  \begin{tabular}{c|cccc}
    \Xhline{1.0pt}
    Methods & NMSE & RMSE & SSIM & PSNR \\
    \hline
    w/o ASPP & 0.0049 & 0.0163 & 0.9485 & 36.53 \\
    w/o residual blocks & 0.0060 & 0.0180 & 0.9431 & 35.58 \\
    w/o height embedding & 0.0072 & 0.0167 & 0.9385 & 36.04 \\
    \hline
    R$^{2}$Net-Out & \textbf{0.0046} & \textbf{0.0156} & \textbf{0.9491} & \textbf{37.00} \\
    \Xhline{1.0pt}
  \end{tabular}
  \label{Ablation outdoor}
\end{table}

\section{Conclusions}
\label{Conclusion}
This paper investigated 3D radio map estimation by exploiting the impact of object height. We first proposed a 2D height embedding method to incorporate height information into 2D images, based on which, a 3D indoor radio map dataset, referred to as 3DiRM3200, was created, which includes a total of $3,200$ 3D radio maps for $200$ buildings. More importantly, the proposed 2D height embedding method enables the design of a 2D deep learning approach for 3D radio map estimation. Specifically, we proposed a 2D deep residual learning method, R$^{2}$Net-In, for indoor 3D radio map estimation by improving the generalization ability of the model and better extracting the features of penetration loss. For outdoor scenarios, R$^{2}$Net-Out was proposed to enhance the feature extraction of diffraction loss, along with a light variant, R$^{2}$Net-Outlite, proposed for the cases with a small training dataset. Quantitative and qualitative experimental results show that compared to the state-of-the-art benchmarks, the proposed R$^{2}$Net-In estimates 3D indoor radio maps more accurately and faster while incurring lower computational and storage costs. For outdoor radio map estimation, the proposed R$^{2}$Net-Out achieves the highest accuracy, while R$^{2}$Net-Outlite surpasses the benchmarks in accuracy, generalization ability, computational cost and memory storage.

Note that due to the limited capability of WinProp, our 3DiRM3200 dataset cannot capture the impacts of dynamic environments, mobility or weather. In fact, the proposed R$^{2}$Net could be further fine-tuned by these available real-world data, which should be carefully studied in the future. In addition, pathloss in other propagation environments, such as rural areas with rich scatterings caused by rugged terrain and mixed environments of indoor and outdoor scenarios, may exhibit significantly different radio propagation characteristics from the indoor and outdoor scenarios studied in this paper. How to customize the proposed R$^{2}$Net to estimate radio maps in rural and mixed environments is an interesting problem, which deserves much attention in future work.

{\small
\bibliographystyle{IEEEtran}
\bibliography{radiomap}
}

\begin{IEEEbiography}[{\includegraphics[width=1in,height=1.25in,clip,keepaspectratio]{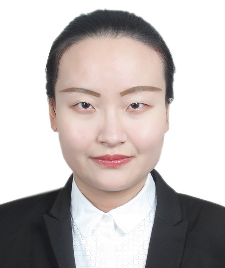}}]{Huiting Rao} 
received the B.S. and M.S. degree in communications engineering from Xiamen University, Xiamen, China, in 2016 and 2019, respecitvely. She is currently pursuing the Ph.D. degree with the College of Electronic and Information Engineering, Tongji University, Shanghai, China. Her research interests include radio map construction, and artificial intelligence.
\end{IEEEbiography}
\vspace{1.5cm}

\begin{IEEEbiography}[{\includegraphics[width=1in,height=1.25in,clip,keepaspectratio]{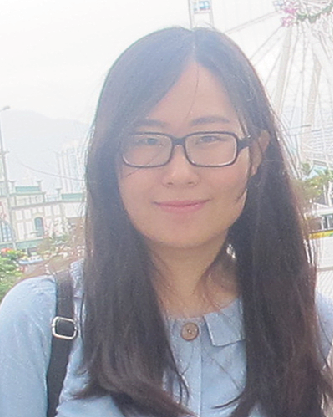}}]{Junyuan Wang} 
(Member, IEEE) received the B.S. degree in communications engineering from Xidian University, Xi’an, China, in 2010, and the Ph.D. degree in electronic engineering from City University of Hong Kong, Hong Kong, China, in 2015. From 2015 to 2017, she was a Research Associate at the School of Engineering and Digital Arts, University of Kent, Canterbury, U.K. From 2018 to 2020, she was a Lecturer (an Assistant Professor) at the Department of Computer Science, Edge Hill University, Ormskirk, U.K. She is currently a Research Professor with the College of Electronic and Information Engineering and the Institute for Advanced Study, Tongji University, Shanghai, China. Her research interests include wireless communications and networking, and artificial intelligence. She was a co-recipient of the Best Paper Award from the IEEE International Conference on Communications in China (ICCC) in 2024 and a co-recipient of the Best Student Paper Award from the IEEE 85th Vehicular Technology Conference–Spring (VTC-Spring) in 2017.
\end{IEEEbiography}

\vspace{-1cm}
\begin{IEEEbiography}[{\includegraphics[width=1in,height=1.25in,clip,keepaspectratio]{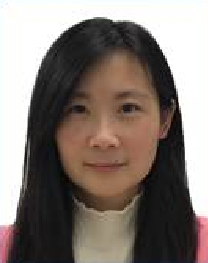}}]{Huiling Zhu} 
(Senior Member, IEEE) received the B.S degree from Xidian University, China, and the Ph.D. degree from Tsinghua University, China. She is currently a Reader in the School of Engineering, University of Kent, United Kingdom. Her research interests are in the area of wireless communications. She was holding European Commission Marie Curie Fellowship from 2014 to 2016. She was Symposium Co-Chair for IEEE Globecom 2015 and IEEE ICC 2018, TPC Co-Chair for IEEE Globecom2026, and Track Co-Chair of IEEE VTC2016-Spring and VTC2018-Spring. Currently, she serves as an Editor for IEEE Transactions on Vehicular Technology.
\end{IEEEbiography}

\vspace{-1cm}
\begin{IEEEbiography}[{\includegraphics[width=1in,height=1.25in,clip,keepaspectratio]{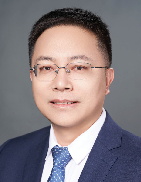}}]{Cheng-Xiang Wang} (Fellow,~IEEE) received the B.Sc. and M.Eng. degrees in communication and information systems from Shandong University, China, in 1997 and 2000, respectively, and the Ph.D. degree in wireless communications from Aalborg University, Denmark, in 2004.
		
He was a Research Assistant with Hamburg University of Technology, Hamburg, Germany, from 2000 to 2001, a Visiting Researcher with Siemens AG Mobile Phones, Munich, Germany, in 2004, and a Research Fellow with the University of Agder, Grimstad, Norway, from 2001 to 2005. He was with Heriot-Watt University, Edinburgh, U.K., from 2005 to 2018, where he was promoted to a Professor in 2011. He has been with Southeast University, Nanjing, China, as a Professor since 2018. He worked as the Dean of the School of Information Science and Engineering, Southeast University from 2020 to 2026, and is now a Vice President of Southeast University. He is also a Professor with the Purple Mountain Laboratories, Nanjing. He has authored four books, three book chapters, and over 720 papers in refereed journals and conference proceedings, including over 280 IEEE journals/magazine papers and 33 highly cited papers. He has also delivered 39 invited keynote speeches and 24 tutorials in international conferences. His current research interests include wireless channel measurements and modeling, 6G/B6G ubiquitous intelligent connectivity networks, and electromagnetic information theory.
		
Dr. Wang is a member of the Academia Europaea (The Academy of Europe) and a member of the European Academy of Sciences and Arts (EASA); a fellow of the Royal Society of Edinburgh (FRSE), IEEE, IET, and China Institute of Communications (CIC); an IEEE Communications Society Distinguished Lecturer in 2019 and 2020; and a Highly-Cited Researcher recognized by Clarivate Analytics in 2017 and 2020 and 2025. He was an Executive Editorial Committee Member of IEEE Transactions on Wireless Communications from 2019 to 2025. He has served as an Editor for over 16 international journals, including IEEE Transactions on Wireless Communications from 2007 to 2009, IEEE Transactions on Vehicular Technology from 2011 to 2017, and IEEE Transactions on Communications from 2015 to 2017. He was a Guest Editor of the IEEE Journal on Selected Areas in Communications, IEEE Transactions on Big Data, and IEEE Transactions on Cognitive Communications and Networking. He has served as a TPC Chair and General Chair for more than 30 international conferences. He received IEEE Neal Shepherd Memorial Best Propagation Paper Award in 2024. He also received 20 Best Paper Awards from international conferences.

\end{IEEEbiography}

\end{document}